\documentclass[12pt,preprint,longabstract]{aastex}
\usepackage{lscape}

\def\msun{$M_{\odot}$}
\def\rsun{$R_{\odot}$}

\def\ergsec{\hbox{erg s$^{-1}$ }}

\begin{document}

\title{A New Dynamical Model for the Black Hole Binary LMC X-1$^{\dagger}$}

\author{Jerome A. Orosz}
\affil{Department of Astronomy, San Diego State University,
5500 Campanile Drive, San Diego, CA 92182-1221}
\email{orosz@sciences.sdsu.edu}

\author{Danny Steeghs}
\affil{Department of Physics, University of Warwick, Coventry, CV4 7AL, UK
and
Harvard-Smithsonian Center for Astrophysics, 60 Garden Street, Cambridge, MA
02138}
\email{D.T.H.Steeghs@warwick.ac.uk}

\author{Jeffrey E. McClintock, Manuel A. P. Torres, Ivan Bochkov,
Lijun Gou, Ramesh Narayan}
\affil{Harvard-Smithsonian Center for Astrophysics, 60 Garden Street,
Cambridge, MA 02138}
\email{jem@cfa.harvard.edu, mtorres@cfa.harvard.edu,
ibochkov@cfa.harvard.edu,lgou@cfa.harvard.edu,
narayan@cfa.harvard.edu}

\author{Michael Blaschak}
\affil{Physics Department, California Polytechnic University, Pomona,
3801 W.\ Temple Avenue
Pomona, CA 91768}
\email{mgblaschak@csupomona.edu}

\author{Alan M. Levine, Ronald A. Remillard}
\affil{Kavli Institute for Astrophysics and Space Research, 
Massachusetts Institute of Technology,
Cambridge, MA 02139-4307}
\email{aml@space.mit.edu, rr@space.mit.edu}

\author{Charles D. Bailyn, Morgan M. Dwyer, Michelle Buxton}
\affil{Department of Astronomy, Yale University, P.O. Box 208101,
New Haven, CT 06520}
\email{charles.bailyn@yale.edu, morgan.dwyer@aya.yale.edu, 
michelle.buxton@yale.edu}

\altaffiltext{$\dagger$}{Based on observations made with
the Magellan 6.5m Clay
telescope at Las Campanas Observatory of the Carnegie Institution}

\begin{abstract}
We present a dynamical model of the high mass X-ray binary LMC X-1 based
on high-resolution optical spectroscopy and extensive optical and
near-infrared photometry.  From our new optical data we find an orbital
period of $P=3.90917 \pm 0.00005$ days.  We present a refined analysis
of the All Sky Monitor data from {\em RXTE} and find an X-ray period of
$P=3.9094 \pm 0.0008$ days, which is consistent with the optical period.
A simple model of Thomson scattering in the stellar wind can
account for
the modulation seen in the X-ray light curves.  The $V-K$ color of the
star ($1.17\pm 0.05$) implies $A_V=2.28\pm 0.06$, which is much larger
than previously assumed.  For the secondary star, we measure a radius of
$R_2=17.0 \pm 0.8 \,R_{\odot}$ and a projected rotational velocity of
$V_{\rm rot}\sin i= 129.9 \pm 2.2$ km s$^{-1}$.  
Using these measured properties to constrain the dynamical model, we
find an inclination of $i=36.38
\pm 1.92^{\circ}$, a secondary star mass of $M_2=31.79\pm
3.48\,M_{\odot}$, and a black hole mass of $10.91\pm 1.41\,M_{\odot}$.
The present location of the secondary star in a temperature-luminosity
diagram is consistent with that of a star with an initial mass of
$35\,M_{\odot}$ that is 5 Myr past the zero-age main sequence.  The star
nearly fills its Roche lobe ($\approx 90\%$ or more), and owing to the
rapid change in radius with time in its present evolutionary state, it
will encounter its Roche lobe and begin rapid and
possibly unstable mass transfer on a
timescale of a few hundred thousand years.
\end{abstract}

\section{Introduction}

The first X-ray source to be discovered in the Magellanic Clouds,
LMC~X-1 (Mark et al.\ 1969), is a persistently luminous ($L_{\rm x} >
10^{38}$~\ergsec) X-ray binary that has been observed by nearly all
X-ray missions during the past 37 years (e.g., Leong et al.\ 1971; Cui
et al.\ 2002).  Spectroscopic studies of its optical counterpart
revealed an orbital period of $\approx 4$ days and a probable mass for
the compact star ``near $M \approx 6$~\msun'' (Hutchings et al.\ 1983,
1987), making it the fourth dynamical black-hole candidate to be
established.

The candidacy of this black hole has had a checkered history.  Even
the most fundamental property of this binary system -- its orbital
period -- was established only very recently.  For nearly
20 years, the
accepted orbital period was 4.2288 d, the value adopted in the dynamical
study of Hutchings et al.\ (1987).  The correct orbital period, $P =
3.9081 \pm 0.0015$~d, was determined by Levine \& Corbet (2006) using
{\it RXTE} All-Sky Monitor (ASM)
X-ray data.  The optical results reported in
this paper and further analysis of the ASM data
amply confirm that the Levine \& Corbet (2006) 
period is correct.  As an
interesting side note, the correct period is noted in the earlier
Hutchings et al.\ (1983) work as one of several candidate orbital
periods.

\begin{figure}
\includegraphics[scale=.85,angle=-90]{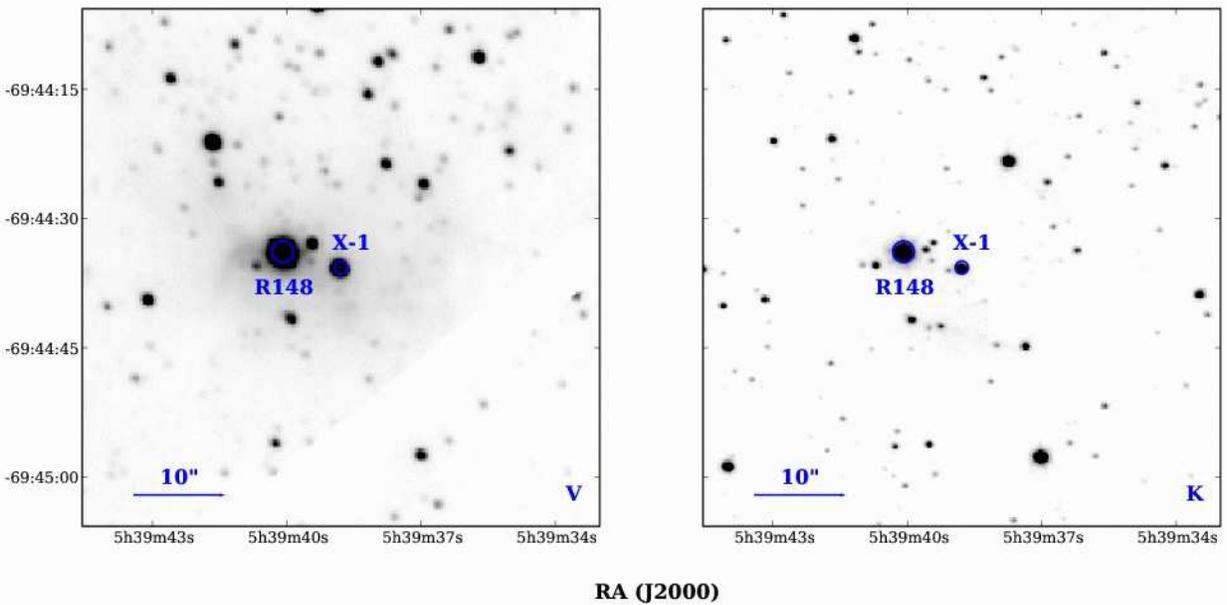}
\caption{$1\times 1$ arcminute finding 
charts for the field surrounding LMC X-1 obtained with the
Magellan-Baade telescope. The left panel is a $V$-band 
image obtained with the
IMACS imaging spectrograph, sampled at 
0\farcs 22 pixel$^{-1}$ under seeing of 0\farcs 8.  The
right panel is the same field observed in the 
$K$-band using the PANIC camera
with 
0\farcs 125 pixel$^{-1}$ 
sampling and a seeing of 0\farcs 55. The correct counterpart
together with the nearby B5 supergiant R148 are marked.}
\label{fc}
\end{figure}

Even the identity of the optical counterpart was not firmly established
until fairly recently.  Initially, based on a rather uncertain X-ray
position, a B5 supergiant known as
R148 was favored over what is now known to be
the counterpart, an O7/O8 giant identified as ``star 32'' by Cowley,
Crampton \& Hutchings
(1978).  
The counterpart was finally established through an analysis of
multiple ${\it ROSAT}$ HRI observations by Cowley et al.\ (1995).  A
precise {\it Chandra} X-ray position (Cui et al.\ 2002) and the
agreement between the X-ray and optical periods (mentioned above) leave
no doubt that star 32 of Cowley et al.\ is the counterpart of LMC X-1.
We present high resolution
$V$- and $K$-band finding charts of the field in Figure
\ref{fc}.

In this paper, we confirm the basic model and the principal conclusions
presented by Hutchings et al.\ (1987) while greatly improving upon their
pioneering work.  Thirty high resolution and
14 medium-resolution spectra have allowed us to
reduce the uncertainty in the radial velocity amplitude $K$ by a factor
of 6 and to obtain a secure value for the rotational line broadening.
Of equal importance, we present the first optical light
curves and infrared magnitudes and colors
of LMC X-1, the analysis of which allows us to strongly constrain
the orbital inclination angle and hence our dynamical model of the
binary.
In \S\ref{photsec} and \S\ref{specsec} we present our new photometric
and spectroscopic observations.  In \S\ref{resultsec}
we present improved measurements of the orbital period, the
temperature of the secondary star, and the radius of the secondary star.
These new observations are used to construct a dynamical model
of the system, which is presented in
\S\ref{discuss}; a summary of our results is in \S\ref{summary}. 

\section{Photometry}\label{photsec}

\subsection{Observations and Reductions}

Optical and near infrared observations of LMC X-1 were obtained on 66
nights between 2007 January 6 and 2007 April 2 using the
ANDICAM\footnote{See http://www.astronomy.ohio-state.edu/ANDICAM.}
instrument on the 1.3m telescope at CTIO, which is operated by the
Small and Moderate Aperture Research Telescope System (SMARTS)
consortium\footnote{See http://www.astro.yale.edu/smarts.}.  During
each visit to the source, a 120 second exposure in the $B$ filter and
a 120 second exposure in the $V$ filter were obtained simultaneously
with an exposure in $J$ that consisted of 10 dithered subexposures of
12 seconds each.  Flat field images for all filters were taken
nightly, and  IRAF tasks were used to perform the standard
image reductions (i.e.\ bias subtraction and flat-fielding for the
optical and flat-fielding, sky subtraction, and image combination for
the infrared).  All of the images were inspected visually, and a few
of them were discarded owing to poor signal-to-noise.  We
retained 58 images in $B$, 57 images in $V$, and 57 images in $J$.

We also observed the field of LMC X-1 on the nights of 2006 December 6
and December 8 using the 6.5m Magellan Baade telescope at Las Campanas
Observatory (LCO)
and the Persson's Auxiliary Nasmyth Infrared Camera
(PANIC; Martini et al.\ 2004).  On December 6 three composite frames
were obtained in the $J$ band and four composite frames were obtained
in the $K$ band in 0\farcs 7 seeing.  Each composite frame consists of
five 7-second dithered subexposures.  On December 8, ten sequences in
$J$ with 3-second subexposures and eight sequences in $K$ with
7-second subexposures were obtained in 0\farcs 5 seeing.  The
conditions were photometric on both nights.  The PANIC data were
reduced and processed with IRAF and custom PANIC software to produce
mosaic frames of the field (13 mosaic frames in $J$ and 12 mosaic
frames in $K$).

\subsection{Derivation of the Photometric Light Curves}\label{photanal}

PSF-fitting photometry was used on the PANIC mosaic images to obtain
instrumental magnitudes of LMC X-1 and two nearby stars.  The absolute
calibration was done with respect to stars from the 2MASS catalog
(Skrutskie et al.\ 2006).  Comparison stars having less than 8000
detector counts in the PANIC frames were selected and checked for
variability. Weighted differential photometry of LMC X-1 was performed
with respect to seven and six comparison stars for the $J$ and
$K$-band mosaic images, respectively.  For LMC X-1, we found a mean
$J$ magnitude of $J=13.76$ with an rms of 0.02 mag and a mean $K$
magnitude of $K=13.43$ with an rms of 0.01 mag.  Optimal aperture
photometry was also performed for LMC X-1 and the comparison stars,
and we obtained similar values for the mean magnitudes.  For
comparison, the magnitudes for LMC X-1 given in the 2MASS catalog are
$J=13.695\pm 0.063$ and $K=13.293\pm 0.063$, although these values
should be treated with caution owing to the presence of the bright
star R148 $\approx 6^{\prime\prime}$ to the east
(Figure \ref{fc}).

The photometric time series for the $B$ and $V$ filters were obtained
using the programs DAOPHOT IIe, ALLSTAR, and DAOMASTER (Stetson 1987,
1992a, 1992b; Stetson, Davis, \& Crabtree 1991).  The instrumental
magnitudes were placed on the standard scales using observations of
the Landolt (1992) fields RU 149, PG 1047+003, and PG 1657+078.
Aperture photometry was done on the standard stars and on several
bright and isolated stars in the LMC X-1 field and the DAOGROW
algorithm (Stetson 1990) was used to obtain optimal magnitudes.  We
find for LMC X-1 an average $V$ magnitude of $V=14.60\pm 0.02$, and a
$B-V$ color of $B-V=0.17\pm 0.08$, marginally consistent with the
values of $V=14.52 \pm 0.05$ and $B-V=0.29\pm 0.02$ given in Bianchi
\& Pakull (1985).

The photometric time series for the SMARTS $J$ band was obtained using
aperture photometry. A relatively small aperture radius ($\approx
1''$) was used to exclude light from nearby sources.  Stars from the
2MASS catalog were used to place the instrumental magnitudes onto the
standard system.  The $J$-band magnitudes from SMARTS are in agreement
with the mean $J$-band magnitude derived from the PANIC data.

\begin{figure}
\epsscale{0.95}
\plotone{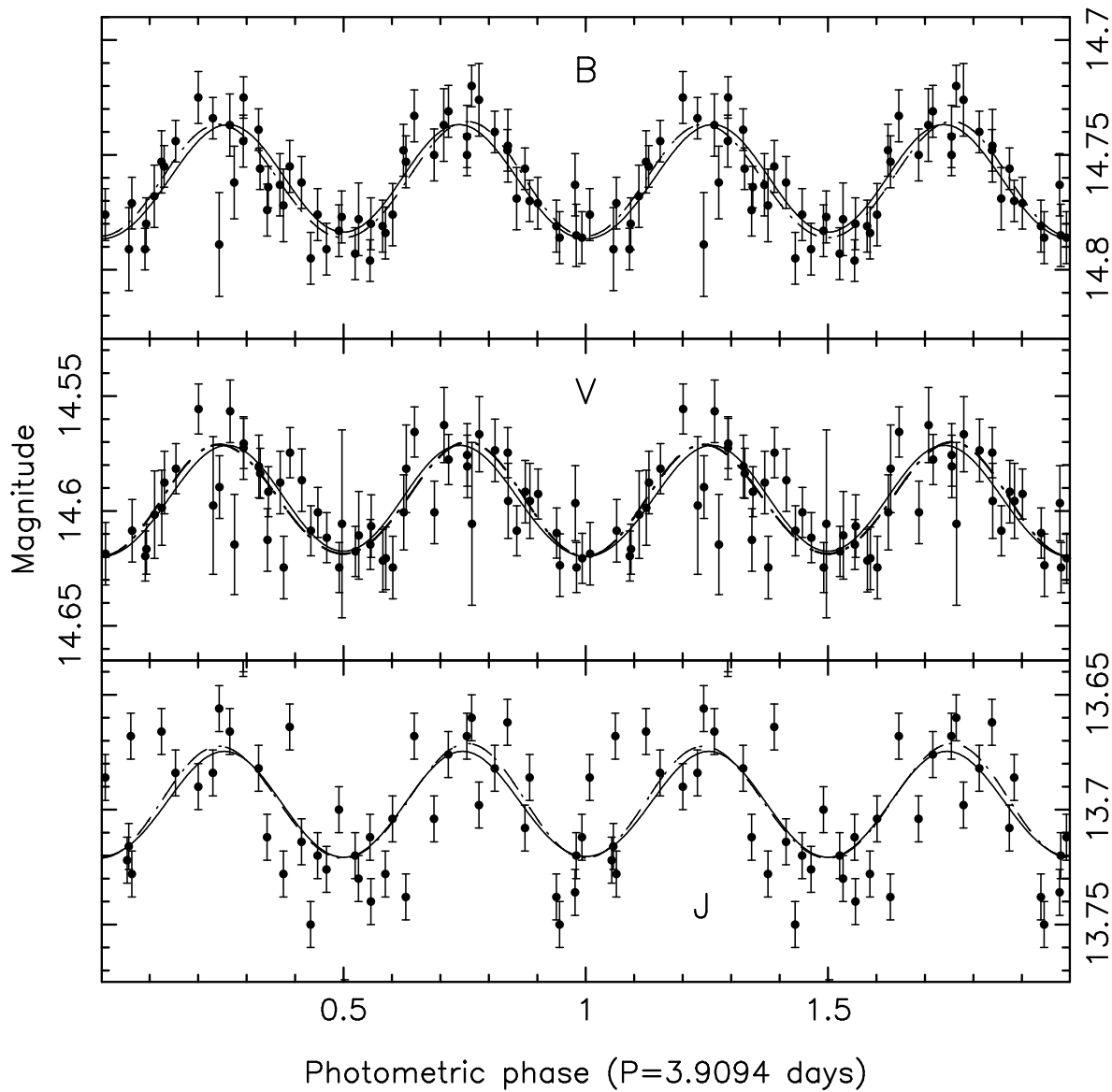}
\caption{The phased light curves of LMC X-1.  Phase zero corresponds
to the time of the inferior conjunction of the secondary star.  Shown
from the top are the $B$, $V$, $J$ light curves with the best-fitting
ellipsoidal models assuming a circular orbit (solid lines) and
assuming an eccentric orbit (dashed lines).  Owing to the large scatter,
the $J$ band data were not used in the modeling.}
\label{lcfig1}
\end{figure}

Figure \ref{lcfig1} shows the light curves phased on the photometric
ephemeris determined below.  The light curves show the 
double-wave modulation characteristic of ellipsoidal variations, with
maxima at the quadrature phases and minima at the conjunction phases.
The amplitude of the modulation ($\approx 0.06$ mag,
maximum to minimum) is not especially
large, which usually indicates a relatively low inclination, or a small
Roche lobe filling factor for the mass donor, or both.

\section{Spectroscopy}\label{specsec}

\subsection{Observations and Reductions}

Thirty
spectra of LMC X-1 were obtained on the nights of 2005
January 19--24 using the Magellan Inamori Kyocera Echelle (MIKE)
spectrograph (Bernstein et al.\ 2002) and the 6.5 m Magellan Clay
telescope at LCO.  The instrument was used in the
standard dual-beam mode with a $1\farcs 0\times5\farcs 0$ slit.
Exposure times ranged between 1200 and 2700 seconds and observing
conditions were good. The seeing was well below $1^{\prime\prime}$
most of the time with excursions as low as 0\farcs 5 and as high as
1\farcs 2.
The pair of $2048\times4096$ pixel CCD detectors were
operated in the $2\times2$ on-chip binning mode.
The blue arm had a
wavelength coverage of 
3340--5065 \AA\ and the spectral dispersion on
the MIT detector was $\lambda/\Delta\lambda=100,000$ (0.03-0.05 \AA\
pixel$^{-1}$), whereas the red arm covered 4855--9420 \AA\ and the
dispersion on its SITe detector was $\lambda/\Delta\lambda=71,000$
(0.07-0.13 \AA\ pixel$^{-1}$). Our $1^{\prime\prime}$ slit delivered a
spectral resolution of $R=33,000$ and 28,000 in the blue and red arms
respectively. ThAr lamp exposures were obtained before and after each
pair of observations of the object, and several flux standards and
spectral-comparison stars were observed.

MIKE was located at a Nasmyth focus, which minimizes flexure and
calibration problems.  No instrument rotator or dispersion
compensation optics were available, and the position angle of the
spectrograph was set to the parallactic angle for an object at air
mass 1.3.  Because all of our observations were taken below air mass
$\approx 1.5$, the effects of light loss due to atmospheric dispersion
are small and unlikely to significantly affect our results.

The MIKE
data were reduced using a pipeline written by
Dan Kelson\footnote{http://www.ociw.edu/Code}.
The
pipeline performs all of the detector calibrations in the standard way
using bias frames and tungsten flat fields which were obtained each
afternoon.  The arc spectra taken at the position of each target were
used to correct for the non-orthogonality between the dispersion and
spatial axes and construct a 2D wavelength solution for all object
frames. All object orders were then sky subtracted
using the technique discussed in Kelson (2003)
and optimally
extracted.
The signal-to-noise ratio per pixel in
the orders with useful lines was generally
in the range of $\approx 50-100$ for most of the spectra.

The extracted spectra were inspected visually and artifacts due to
cosmic rays were removed manually using simple interpolation.  In
addition to cosmic rays, there were some artifacts in the cores of the
higher Balmer lines owing to imperfect subtraction of nebular lines.
These artifacts were also removed by interpolation.

\begin{figure}
\epsscale{0.95}
\plotone{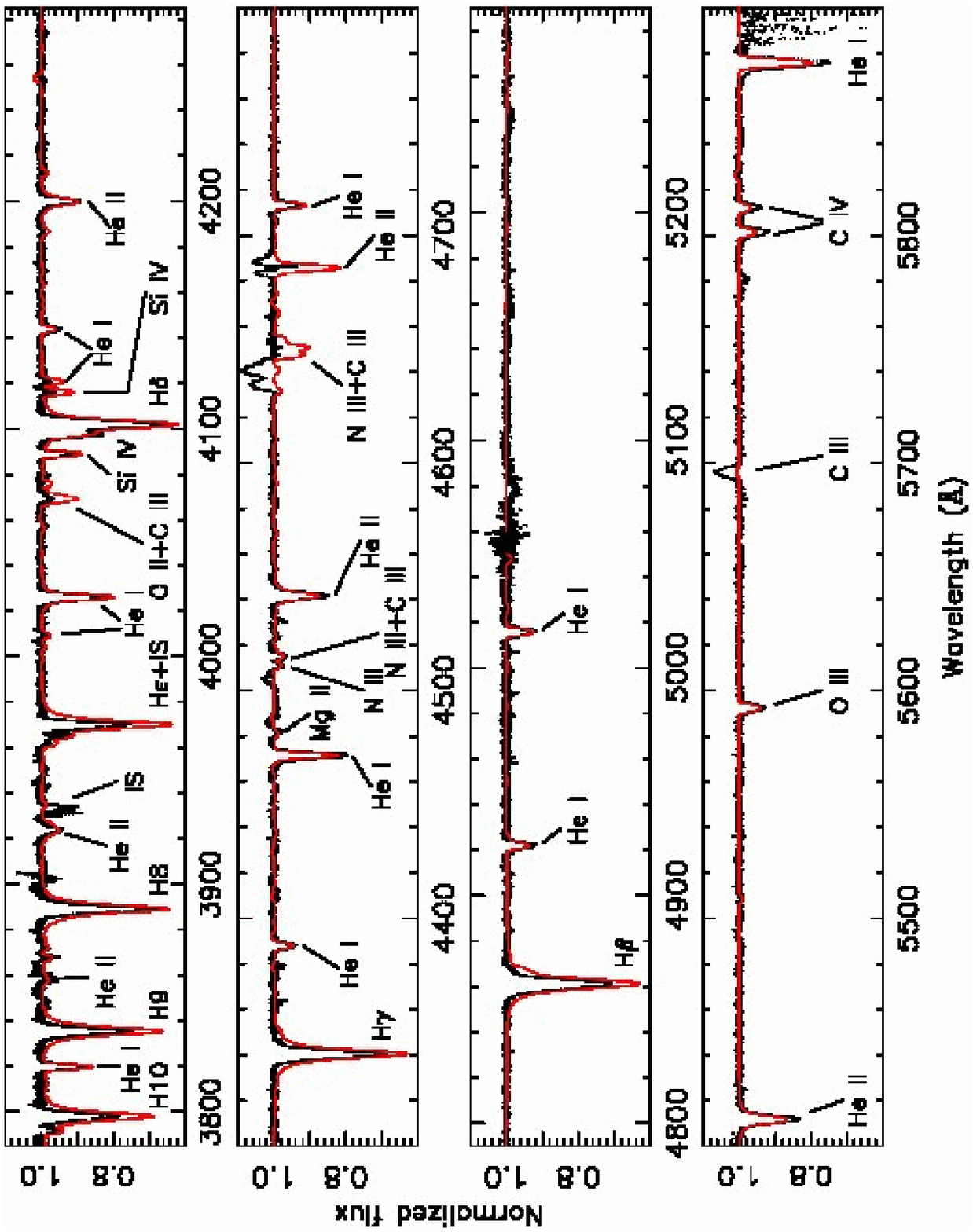}
\caption{The average spectrum of LMC X-1 in the restframe of the
secondary star (black dots) obtained with the MIKE echelle.
Most of the strong features are labeled, and interstellar
lines are marked with `IS'. 
The best-fitting model (red line) has
$T_{\rm eff}=33,225$~K, $\log g=3.56$, and $V_{\rm rot}\sin i
= 128.0$ km s$^{-1}$.  The noise near 5060~\AA\
is caused by the transition from the blue arm to the red arm.
}
\label{newspecfig}
\end{figure}

Finally,
the individual orders in the cleaned spectra
were trimmed to remove the low signal-to-noise regions
at each end
and normalized using cubic spline fits.
The  trimmed and normalized
orders were then merged into a pair of spectra (e.g.\
red arm and blue arm).
Figure \ref{newspecfig} shows the average
spectrum (in the restframe of the secondary) for much of the blue arm
and part of the red arm.  The model spectrum shown in
Figure \ref{newspecfig} is discussed in \S\ref{tempsec}.

An additional fourteen spectra of LMC X-1 were obtained on the nights
of 2008 August 2-3 using
the Magellan Echellette Spectrograph (MagE; Marshall et al.\ 2008)
and the Clay telescope at LCO.  We used the 0\farcs 7 slit
which yields a resolving power of $R=6000$.
The exposure times were 600 seconds and the seeing ranged from
0\farcs 8 to 1\farcs 1 for the first night and
1\farcs 4 to 2\farcs 7 for the second night.  ThAr lamp exposures
were obtained before and after each pair of observations
of LMC X-1.

The images were reduced with tasks in the IRAF `ccdproc' and `echelle'
packages.  After the bias was subtracted from each image, pairs of
images were combined using a clipping algorithm to remove cosmic rays.
The resulting seven images were flat-fielded using a normalized master
flat and then rotated in order to align the background night sky
emission lines along the columns.  After the background emission lines
were rectified, the spectra from individual orders were optimally
subtracted.  Unfortunately, the rectification of the background lines
was not perfect, which produced artifacts due to nebular lines in the
cores of the higher Balmer lines.  These imperfections were removed by
simple interpolation.  The process used to merge the orders in the MIKE
spectra described above was also used to merge the MagE orders.  The
resulting seven spectra cover a useful wavelength range of
3200-9250~\AA.

\subsection{Radial Velocities}\label{radvel}

In a recent study of three high mass X-ray binaries with OB-star
companions, van der Meer et al.\ (2007) measured radial velocity curves
using single lines.  They found that X-ray heating produced distortions
in some of the line profiles, thereby yielding for some cases different
$K$-velocities for different lines.

In order to assess the effects of X-ray heating on our spectra, we
selected four bandpasses with strong lines that have a wide range of
excitation energies over which to derive radial velocities from the MIKE
spectra: 3780--3880~\AA, dominated by H9 \& H10 and including He I
$\lambda3820$; 3990-4062~\AA, He I $\lambda4026$; 4150--4250~\AA, He II
$\lambda4200$; and 4425--4525~\AA, He I $\lambda$ 4471.  These lines
were selected because of their prominence in both object and template
spectra and their isolation from other spectral features.

The radial velocities were determined using the  {\sc fxcor}
task within IRAF, which is an implementation of the cross-correlation
analysis developed by  Tonry \& Davis (1979).
For each bandpass, the spectra were low-frequency filtered either by
fitting a Legendre polynomial or by Fourier filtering.  To optimize
the removal of any order blaze remnants, the Legendre polynomial
orders and the Fourier ramp filter components were determined
individually for each bandpass. For example, the region
containing H9 and H10
required a Legendere polynomial order of 8 (i.e.\ nine terms) or a
Fourier cut-on wavenumber of 4 and a full-on wavenumber of 8, whereas
the region containing the line He I $\lambda4471$ needed a 28-piece
polynomial or a cut-on of 9 and a full-on of 18.
The spectra were then continuum-subtracted
prior to computing the cross-correlation.  No other filtering was
applied.

The correlations were performed using three different templates: a
model spectrum from the OSTAR2002 grid (Lanz \& Hubeny 2003;
see
the discussion below) and the spectra of
two of our comparison stars: HD~93843, O6~III(f) and HD~101205,
O7~IIIn((f)) (Walborn 1972, 1973).  Typical values of the Tonry \&
Davis (1979) $r$-value, a measure of signal-to-noise ratio, ranged
from about 10 to 70 and the median value was 26.

The radial velocity data for each bandpass were fitted to a circular
orbit model which returns the systemic velocity $V_{\rm 0}$, the time of
maximum velocity $T_{\rm max}$, and the velocity semiamplitude of the
secondary $K_2$.  The orbital period was fixed at our adopted
value, which is given in Table 3.  The model provided good fits to
the data.  For each fit, the statistical errors on the velocities
returned by the fit were scaled by factors ranging from 0.3 to 2.1 and
the data were refitted in order to give a reduced chi-squared
($\chi_{\nu}^{2}$) of unity.

\begin{figure}
\epsscale{1.05}
\plotone{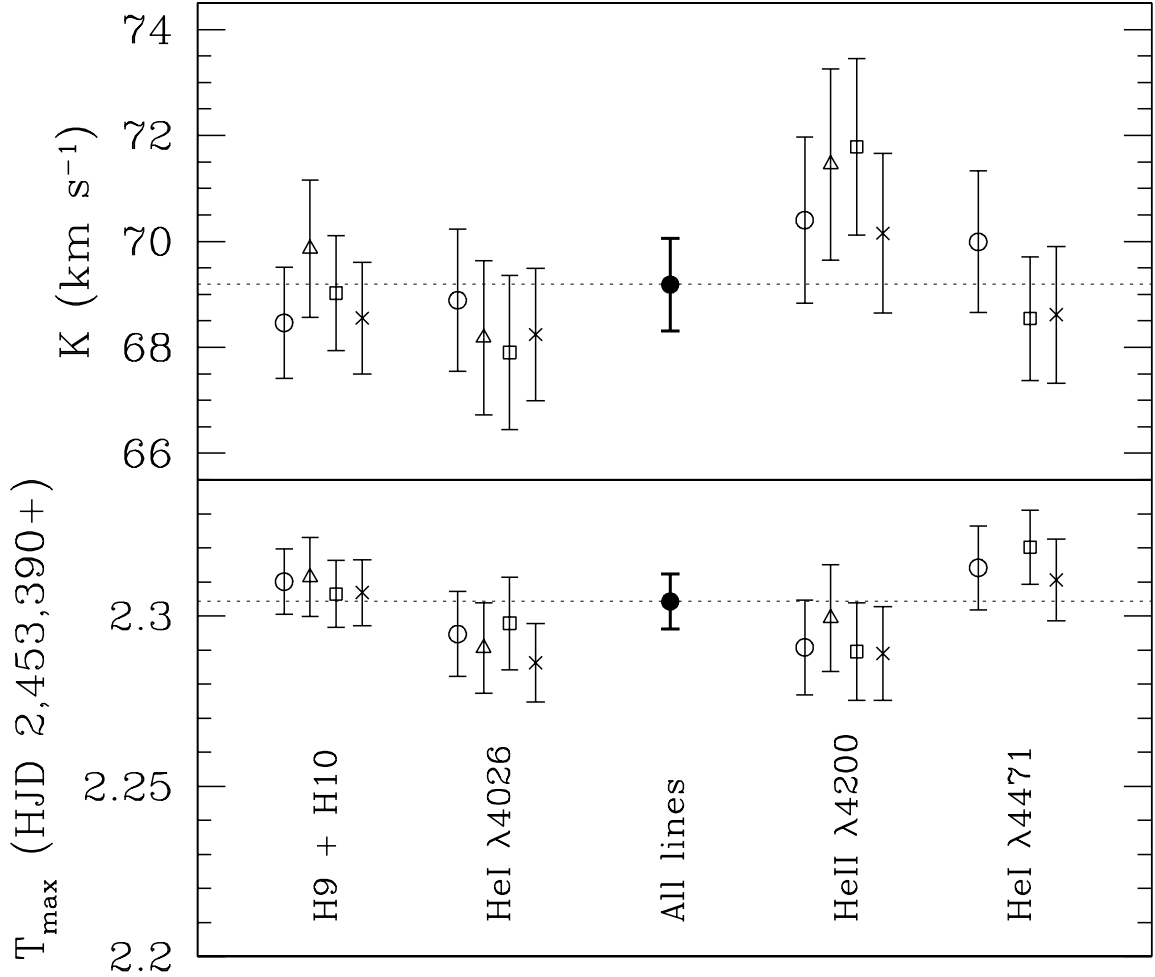}
\caption{Summary of fitted values of the velocity semiamplitude $K$
(top) and the time of maximum velocity $T_{\rm max}$ (bottom) for
different combinations of bandpasses, template spectra and modes of
filtering.  The lines or line group are indicated at the bottom of the
figure (see text).  The open and filled circles are for template HD
93843 and Legendre filtering.  Triangles: HD 101205, Legendre filtering.
Squares: Synthetic template, Legendre filtering.  Crosses: Synthetic
template, Fourier filtering.  The filled circles and dashed lines
correspond to the values of $K$ and $T_{\rm max}$ given in the text,
which were computed for the mean velocities.  The uncertainties shown
are purely statistical, correspond to $\chi_{\nu}^{2}=1$, and are at the
$1\sigma$ level of confidence.  No results are shown for the He I
$\lambda4471$ line for HD 101205 because the template spectrum was
corrupted in this band.}
\label{Kfig}
\end{figure}

As an aid to visualize the systematic effects, Figure \ref{Kfig}
summarizes the fitted values of $K_2$ and $T_{\rm max}$ obtained for 
different combinations of
(i)
the four bandpasses, (ii) the three template stars and (iii) the two
modes of low-frequency filtering, Legendre and Fourier.  The open
circles show, for example, the results obtained by fitting the velocity
data using the spectrum of HD 93843 as the template and employing
Legendre filtering.  The other open symbols show for comparison the
results obtained with the other template/filter combinations.  The
results shown as filled circles and bold error bars were derived using
the weighted mean velocities for each spectrum: $K_2 = 69.19 \pm 0.88$
km s$^{-1}$ and $T_{\rm max} = {\rm HJD~} 2453392.3043 \pm 0.0081$.
These mean values are indicated by the dashed lines.

Figure \ref{Kfig} shows that our results for $K$ and $T_{\rm max}$ are
quite insensitive to the choice of bandpass, template or mode of
filtering, although there are some systematic effects at play.  For
example, the weighted mean $K$-velocity of the He II line, $K
$($\lambda4200$)$ =70.87 \pm 0.81$ km s$^{-1}$, is greater than our
adopted mean value, albeit still within 2$\sigma$.
This modest shift in $K$ is far
less than was observed in the  recent van der Meer
et al.\ (2007) study of three O-star X-ray
binaries that contain neutron star primaries.
Because we have considered lines with a wide
range of excitation energies (Balmer, He I and He II) and have found
quite consistent results, we conclude that our dynamical results are
little affected by X-ray heating, tidal distortion of the secondary,
stellar wind, and other commonly-observed sources of systematic
effects.

\begin{figure}
\epsscale{0.85}
\plotone{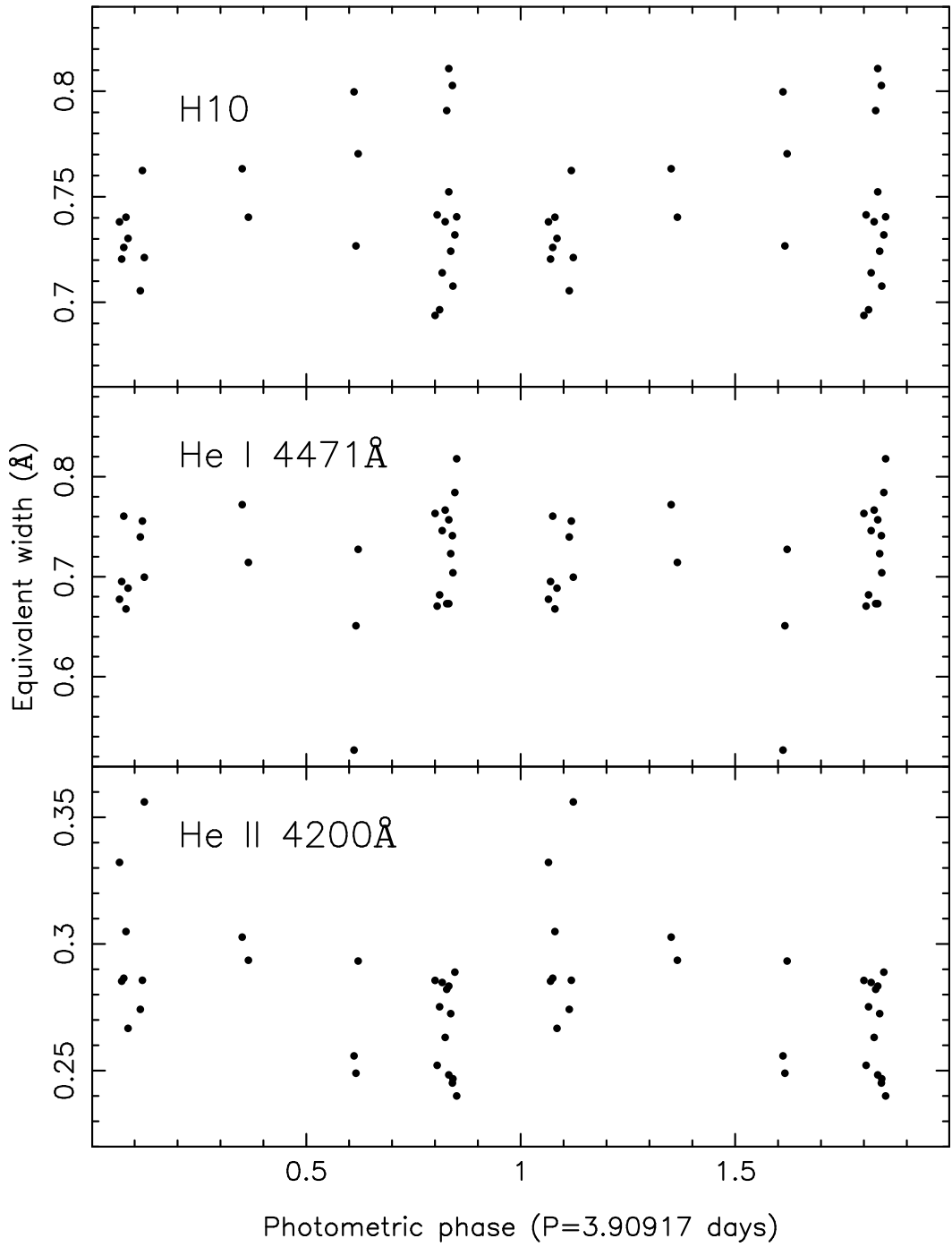}
\caption{The equivalent widths of H10 (top), He I $\lambda 4471$ 
(center), and He II $\lambda 4200$ (bottom) as a function of
orbital phase, where phase zero corresponds
to the time of the inferior conjunction of the secondary star.  
}
\label{ewfig}
\end{figure}

Finally, we measured the equivalent widths of the H10, He I
$\lambda 4471$, and He II $\lambda 4200$ lines in the individual
spectra.  There is no apparent trend with orbital phase
(see Figure \ref{ewfig}), which 
further confirms that the effects of X-ray heating are minimal.

\begin{figure}
\epsscale{0.85}
\plotone{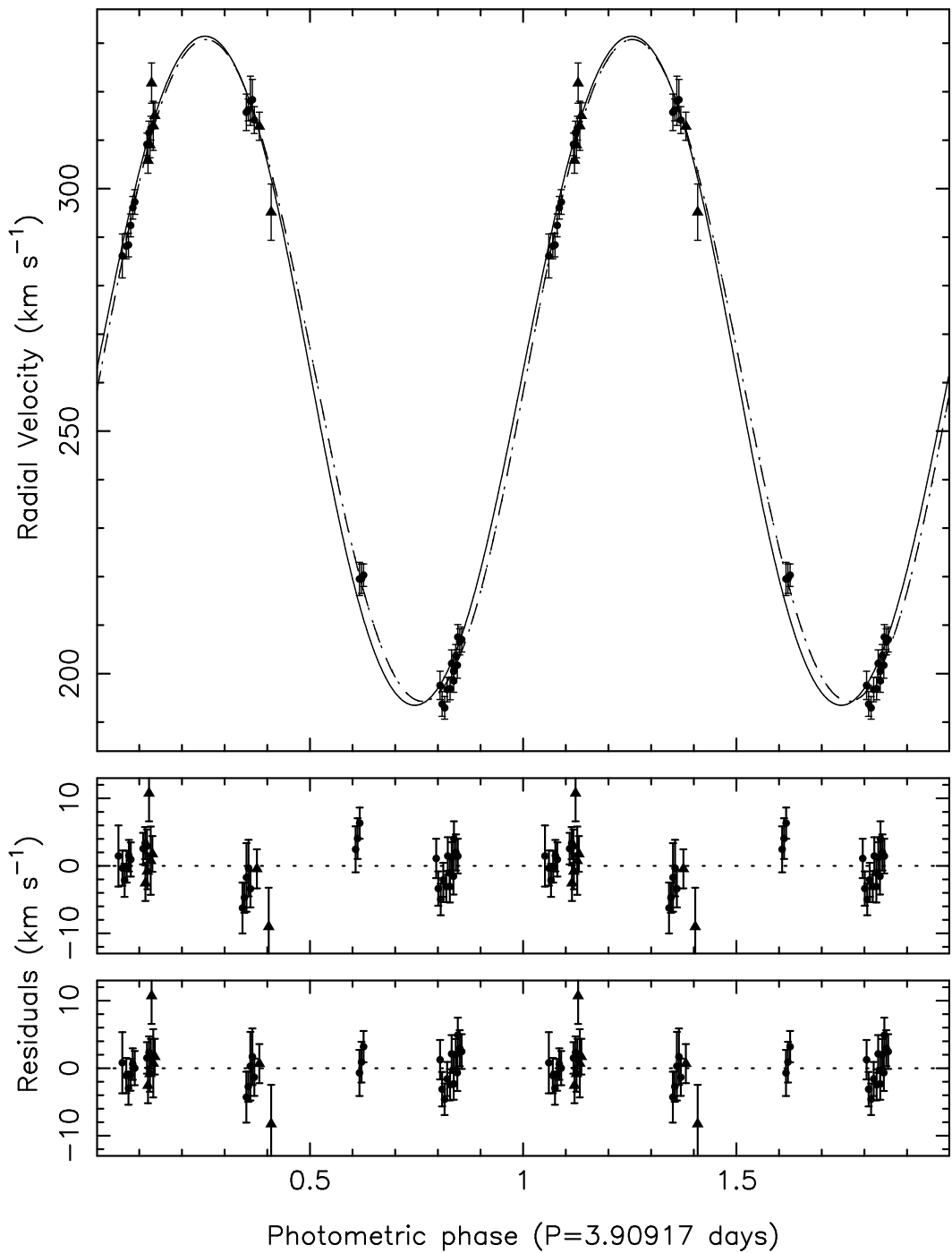}
\caption{Top:
The phased velocity curve of LMC X-1.  Phase zero corresponds
to the time of the inferior conjunction of the secondary star.  
The MIKE radial velocities are shown with the filled circles and the
MagE radial velocities are shown with the filled triangles.
The
model curve for the best-fitting orbital model assuming a circular
orbit is shown with the solid line, and the model curve for the
best-fitting model assuming an eccentric orbit is shown with the dashed
lines.
Center:  The residuals with respect to the circular orbit model.
Bottom:  The residuals with respect to the eccentric orbit model.
}
\label{rvfig1}
\end{figure}

Knowing that the velocities are little effected by X-ray heating, we
derived the final adopted velocities from the MIKE and MagE spectra
using  {\sc fxcor}, a synthetic template,
and a cross correlation region that covers
nine He I and He II lines between 4000~\AA\ and 5020~\AA.
The mean velocity curve is shown plotted in Figure
\ref{rvfig1}.

\section{Results}\label{resultsec}

\subsection{Refined Orbital Period}\label{periodogram}

Levine \& Corbet (2006) analyzed {\em RXTE} ASM data of LMC X-1 taken
between 1996 March and 2005 November.  They reported a periodicity of
$3.9081 \pm 0.0015$ days, which differs from the period of $P=4.2288$
days given in Hutchings et al.\ (1987).  On the other hand, the X-ray
period is consistent with one of the possible periods given in
Hutchings et al.\ (1983), namely $P=3.909 \pm 0.001$ days.

\begin{figure}
\includegraphics[scale=.7]{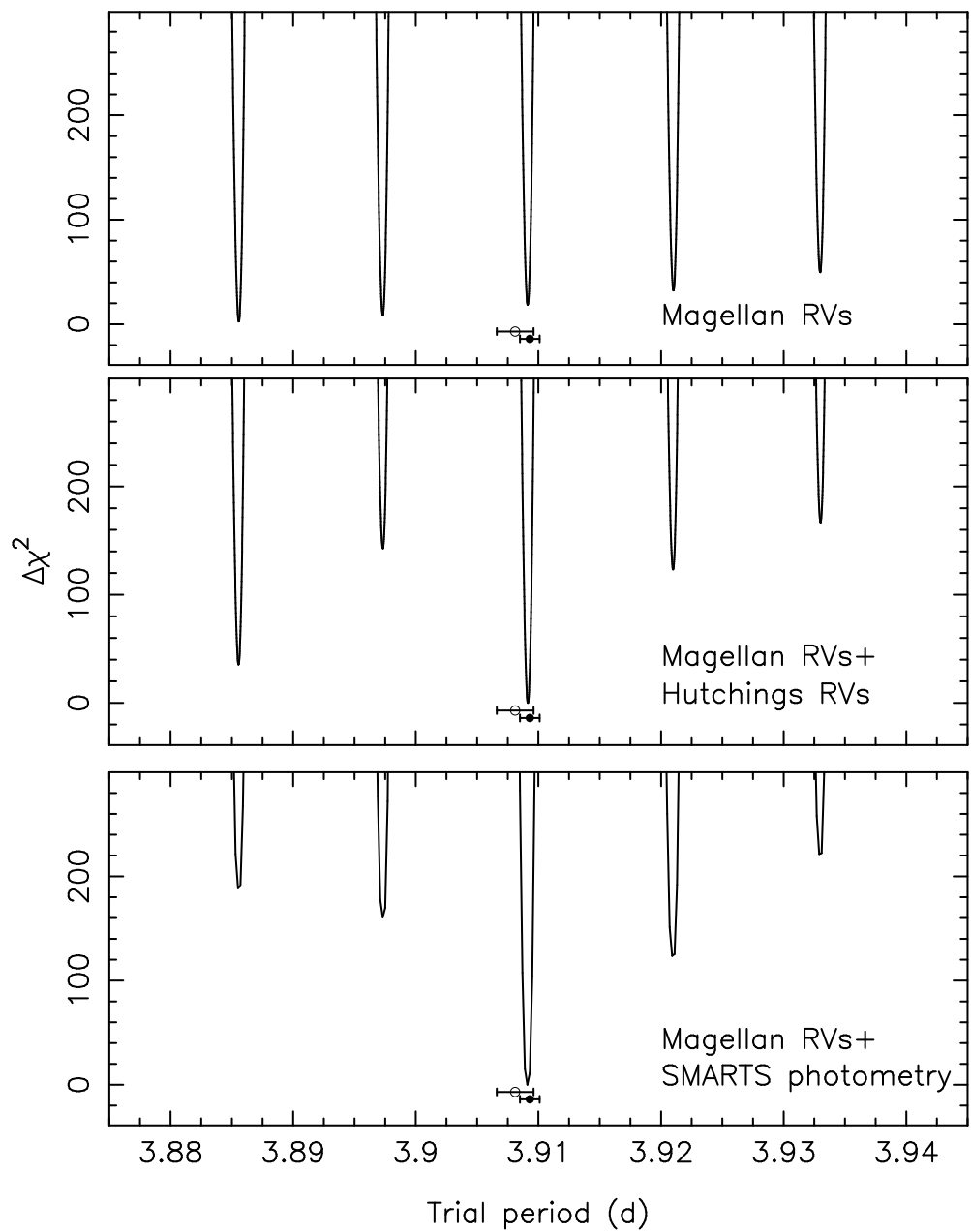}
\caption{Top:  $\chi^2$ computed using a three-parameter
sinusoid vs.\ the trial period for the Magellan radial velocities. 
Center:  The periodogram (computed in a similar manner as 
above) for the  radial velocities from Hutchings et
al.\ (1983, 1987) combined with the Magellan
radial velocities.  
Bottom:  The periodogram derived from the SMARTS photometry and
the Magellan radial velocities (see text).
The best-fitting period is found to
be $3.90917 \pm 0.00005$, and all other possible alias periods are
ruled out at high confidence.
In each of the three panels
the X-ray
period given in Levine \& Corbet (2006) is indicated with an open
circle, and the refined X-ray period is denoted by the filled circle.
}
\label{plotperiodogram}
\end{figure}

We derived a refined period from our optical data.  The optical light
curves completely rule out the 4.2288 day period and strongly favor
the period near 3.909 days.  We made a periodogram  from the
Magellan radial velocities by fitting a three-parameter sinusoid 
to
the data at various trial periods and recording the $\chi^2$ of
the fit.  The results are shown in the top of Figure \ref{plotperiodogram}.
A unique period cannot be found from these data alone, leaving us with
possible periods  near 3.884, 3.898 and 3.909 days.   
Although the radial velocities
given in Hutchings et al.\ (1983, 1987) have relatively large errors
and the times given are accurate to only about one minute,
they can be used to rule out possible alias periods.  When these
additional velocities are included in the analysis, only the period
near 3.909 days remains viable (see
the center panel in Figure \ref{plotperiodogram}).
Finally,
we used the ELC code (Orosz \& Hauschildt
2000) to compute periodograms by modeling both our light and velocity
curves together.  The details of the model fitting are discussed
thoroughly in \S\ref{ELC}.  
For the purposes of computing a periodogram, the
period is held fixed at a given trial period and several other fitting
parameters are varied (here the orbit is assumed to be circular) until
the fit is optimized.  The graph of $\chi^2$ vs.\ $P$ (see the bottom
panel of Figure
\ref{plotperiodogram}) then serves as a periodogram.  The combination
of the SMARTS photometry and the Magellan radial velocities yields 
a unique period of $3.90917\pm 0.00005$ days.
This period is consistent with the period found from the combination
of the Hutchings et al.\ (1983, 1987) and Magellan radial velocities.
This period is also consistent with the X-ray period found by
Levine \& Corbet (2006), and with the refined X-ray period that
we now derive.

The ASM consists of
three Scanning Shadow Cameras (SSCs) mounted on a rotating Drive
Assembly \citep{asm96}.  Approximately 53,700 measurements of the
intensity of LMC X-1, each from a 90-s exposure with a single SSC,
were obtained from the beginning of the {\it RXTE} mission in early
1996 through 2008 June.  A single exposure yields intensity estimates
in each of three spectral bands which nominally correspond to photon
energy ranges of 1.5-3, 3-5, and 5-12 keV with a sensitivity of a few
SSC counts s$^{-1}$ (the Crab Nebula produces intensities of 27, 23,
and 25 SSC counts s$^{-1}$ in the 3 bands, respectively).  The X-ray
intensity of LMC X-1, as seen over more than 12 years in the ASM light
curves, has been more or less steady near 20 mCrab (1.5-12 keV) with
variations, when 10-day averages are considered, of $\approx \pm 10$\%.

A periodicity in the X-ray intensity of LMC X-1 was found during a
search for periodicities in ASM data using advanced analysis
techniques.  Two strategies were applied in this search to improve the
sensitivity.  The first involves the use of appropriate weights such
as the reciprocals of the variances, 
since the individual ASM measurements have a wide range of
associated uncertainties.  The second strategy stems from the fact
that the observations of the source are obtained with a low duty
cycle, i.e., the window function is sparse (and complex).  The
properties of the window function, in combination with the presence of
slow variations of the source intensity act to hinder the detection
of variations on short time scales.  The power density
spectrum (PDS) of the 
window function   has substantial power at high frequencies, e.g., 1
cycle d$^{-1}$ and 1 cycle per spacecraft orbit ($\approx 95$ minute
period).  Since the data may be regarded as the product of a
(hypothetical) continuous set of source intensity measurements with
the window function, a Fourier transform of the data is equivalent to
the convolution of a transform of a continuous set of intensity
measurements with the window function transform.  The high frequency
structure in the window function transform acts to spread power at low
frequencies in the source intensity to high frequencies in the
calculated transform (or, equivalently, the PDS).  This effectively
raises the noise level at high frequencies.

\begin{figure}
\epsscale{0.85}
\plotone{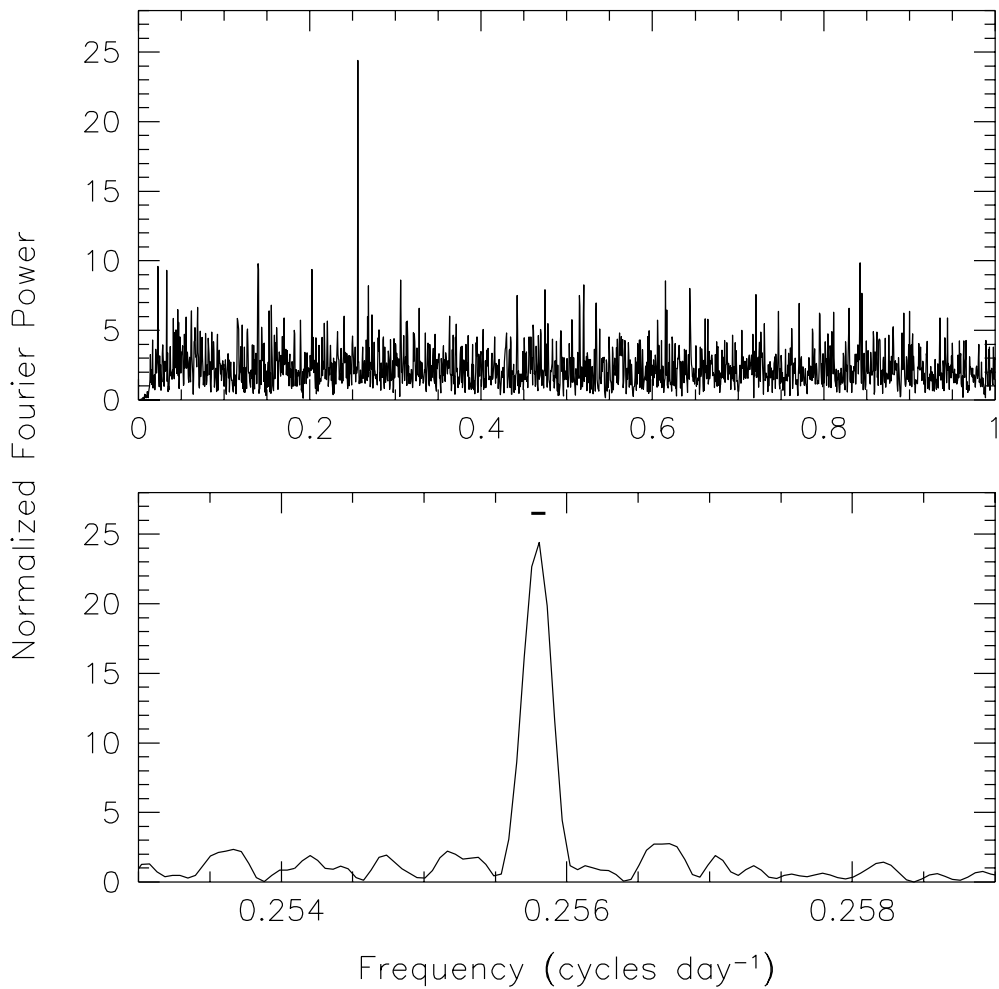}
\caption{Power density spectrum (PDS) of a 1.5-12 keV ASM light curve
of LMC X-1 that has been modified to remove variability on time scales
longer than $\approx 30$ days (see text).  The original FFT was
oversampled by a factor of four and had approximately 62,000
frequencies from 0 cycles day$^{-1}$ to the Nyquist frequency of 3.3
cycles day$^{-1}$.  Top: The low frequency part of a rebinned PDS in
which the number of frequency bins was reduced by a factor of 10 by
using the maximum power in each contiguous set of 10 frequency bins in
the original PDS as the power of the corresponding bin in the rebinned
PDS.  Bottom: A portion of the original PDS. The horizontal bar above
the peak shows our estimate of the possible values of the centroid
frequency, i.e., $0.25580 \pm 0.00005 $ d$^{-1}$.  In both panels, the
power is normalized relative to the PDS-wide average.}
\label{ASM_PDS}
\end{figure}

In our analysis, the sensitivity to high frequency variations is
enhanced by subtracting a smoothed version of the light curve from the
unsmoothed light curve.  To perform the smoothing, we do not simply
convolve a box function with the binned light curve data since that
would not yield any improvement in the noise level at high
frequencies.  Rather, we ignore bins which do not contain any actual
measurements and we use weights based on estimates of the
uncertainties in the individual measurements to compute the smoothed
light curve.  
The kernel used in the smoothing was a Gaussian with a
full width at half maximum of 30.0 days, 
so the smoothed light curve
displayed only that variability with Fourier components at frequencies
below $\approx 0.03$ day$^{-1}$. The smoothed light curve was
subtracted from the unsmoothed light curve, and the difference light
curve was Fourier transformed.  The results are illustrated in Figure
\ref{ASM_PDS}.  The center frequency of the peak and the formal
uncertainty thereof are $0.25580 \pm 0.00005 $ d$^{-1}$.  The
corresponding period is $3.9094 \pm 0.0008$.  This period differs from
the optical period by only $1.3\times 10^{-4}$ days.

In Figure \ref{ASMfold} we show folded X-ray light curves for the
1.5-3, 3-5, and 5-12 keV photon-energy bands as well as for the
overall 1.5-12 keV band.  There is a clear orbital modulation with the
minimum intensity occurring at phase 0, which corresponds to the time
of the inferior conjunction of the secondary star.  The X-ray light
curves can be fitted to a function of the form $f(\phi) = a_0 +
a_1\cos(2\pi\phi) + a_2\sin(2\pi\phi)$.  Table \ref{ASMfits} gives the
best-fitting coefficients for the three individual bands and for the
sum.  In all cases, the $a_2$ terms are consistent with zero.  We
discuss this modulation in more detail in
\S\ref{wind}.

\subsection{Temperature and Rotational Velocity}\label{tempsec}

The effective temperature of the secondary star and its (projected)
rotational velocity are needed as inputs and constraints on the
dynamical model discussed below. Previously, Hutchings et al.\ (1983)
derived a spectral type of O7 for the secondary star based on the line
strengths of He I, He II, Si IV, and Mg II and a rotational velocity
of $V_{\rm rot}\sin i \approx 150$ km s$^{-1}$.  Negueruela \& Coe
(2002) derived a spectral type of O8III based on the ratio of the He
II $\lambda4541$ and He I $\lambda4471$  lines and on the
strength of the Mg II $\lambda4481$  line.  A spectral type of
O8III corresponds to an effective temperature of around 33,000~K
(Heap, Lanz, \& Hubeny 2006).

We used model spectra from the OSTAR2002 grid (Lanz \& Hubeny 2003) to
derive improved values of the temperature and rotational velocity of
the secondary.  The temperature and gravity sampling of the grid was
extended to 50~K and 0.05 dex, respectively, by first resampling the
high dispersion model spectra to a fine wavelength spacing common to
all models and then by interpolation of the resampled models.  Hubeny's
program ROTINS3 was used to broaden model spectra with instrumental
and rotational broadening profiles, and to resample the resulting
models to match the wavelength sampling of the observed spectra.

The ``surface'' gravity of the secondary star is tightly constrained by
the dynamical solution discussed below and is found to be $\log g=3.50$
with a formal $1\sigma$ error less than 0.02 dex.  
For the purposes of finding the temperature we allow $\log g$ to fall
within the generous range of $3.4\le \log g\le 3.6$.
With our large grid of model spectra over a large range of
temperatures, gravities, and rotational velocities, we performed
$\chi^2$ tests using nine helium lines between 4000~\AA\ and 5020~\AA.
The line ratios of He II $\lambda4541$/He I $\lambda 4471$ and He II
$\lambda 4200$/He I(+II) $\lambda4026$, which are the main
classification criteria for O-stars (e.g.\ Walborn \& Fitzpatrick 1990),
are included in this range.  We find a temperature of $T_2=33,225\pm
75$~K from the average MIKE spectrum and $T_2=33,200 \pm 150$~K from the
average MagE
spectrum.  Figure \ref{newspecfig} shows the best-fitting model with the
MIKE spectrum.  Generally the lines in the model spectrum match up with
the lines in the observed spectrum.  There are a few exceptions such as
the O II + C III blend near 4035~\AA, the Bowen blend near 4640~\AA, the
He II line near 4686~\AA, and the C III line near 5690~\AA.  The wings
and the cores of the Balmer lines are not well fit.  However, in the
case of the MIKE spectra it is difficult to normalize the profiles of
the Balmer lines owing to their large widths which can take up a
substantial part of an individual order.  The Balmer lines are matched
better in the MagE spectrum where it is much easier to normalize the
individual orders owing to the lower spectral resolution.

The errors on the temperature quoted above are the formal
$1\sigma$ errors. The true uncertainty on the effective
temperature is no doubt larger owing to the fact that the model
spectrum is not a perfect fit to the data.  Also, there
are almost certainly systematic errors in the model atmospheres
as well.  Accordingly, we will adopt a temperature of
$T_2=33,200$~K with an uncertainty of 500~K
for the analysis presented below.

\begin{figure}
\includegraphics[scale=0.75,angle=-90]{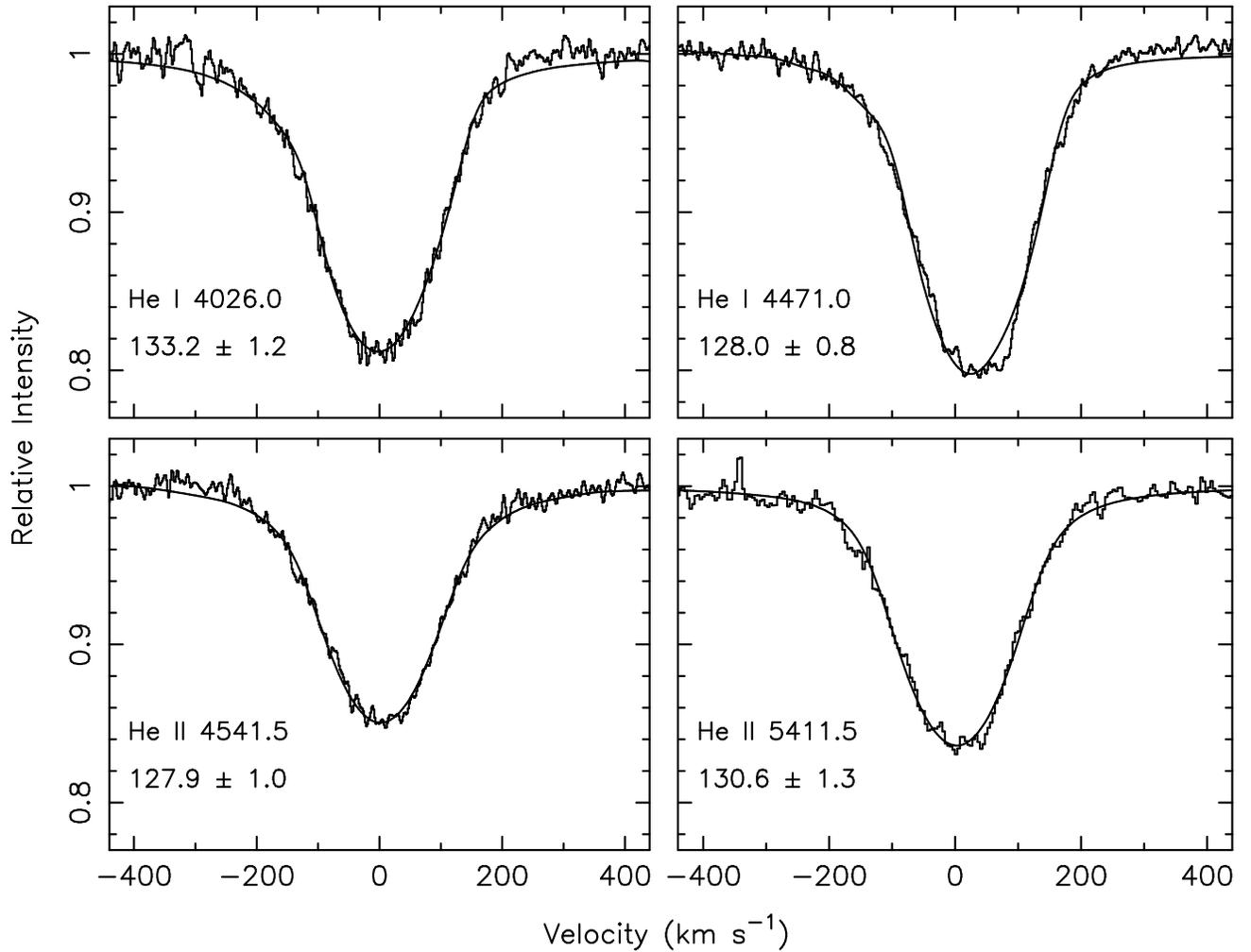}
\caption{The average profiles of the four helium lines used to
determine the rotational velocity (``histograms''), and
the best-fitting model profiles for each (smooth curves).  
The line identification and the
derived value of $V_{\rm rot}\sin  i$ in km s$^{-1}$ are given
in each panel.
}
\label{vrotfig}
\end{figure}

To refine our initial measurement of the rotational velocity of the
star, we used the He I lines near 4026.0~\AA\ and 4471.0~\AA\ and
the He II lines near 4541.4~\AA\ and 5411.4~\AA.
We computed large grids of models
with various temperatures, gravities, and rotational velocities 
with the appropriate resolving power for the MIKE 
spectrograph and
performed $\chi^2$ tests to the composite line profiles (here we
assume that the main broadening mechanism in the line profiles is
rotation).  Figure \ref{vrotfig} shows the results.
The model profiles fit the observed profiles fairly well.
The rotational velocities found from the individual lines range from
$127.9\pm 1.0$ km s$^{-1}$ to $133.2\pm 1.2$
km s$^{-1}$, where the uncertainties
are the formal $1\sigma$ errors.  The average value is
129.9 km s$^{-1}$, and the standard deviation of the four measurements
is 2.18 km s$^{-1}$.  Hence for the analysis below, we adopt a value of
$V_{\rm rot}\sin i=129.9\pm 2.2$ km s$^{-1}$.

\subsection{Radius of the Secondary Star and Reddening}

Unlike the case for most Galactic X-ray binaries, the distance to LMC
X-1 is well-determined.  
Given  the distance, we can find the
radius of the secondary star if we can obtain good measurements of the
apparent magnitude of the star, its effective temperature, the
extinction, and the bolometric correction.  If the radius of the
secondary star can be found independently, then the dynamical model of
the system is further constrained.

The apparent magnitude of the system is easy to measure.  The spectral
type of the star yields its effective temperature, and the bolometric
corrections can be estimated using detailed model atmosphere
computations (e.g.\ Lanz \& Hubeny 2003; Martins \& Plez 2006).  
The remaining quantity, the extinction to the source, is unfortunately
subject to the largest uncertainties and difficult to measure.

Using spectra from the {\em International Ultraviolet Explorer} (IUE),
Bianchi \& Pakull (1985) determined a color excess of $E(B-V)=0.37$
(0.32 from extinction in the LMC and 0.05 Galactic foreground
extinction).  Bianchi \& Pakull (1985) noted that the color excess
derived from the IUE data is lower than the value derived from the
optical color of the star.  Their value of $B-V=0.29$ and
$(B-V)_0=-0.27$ (Martins \& Plez 2006) gives $E(B-V)=0.56$ Using our
measured value of $B-V=0.17$ we find $E(B-V)=0.44$.  Thus we have a
range of color excess values of $0.37 \le E(B-V) \le 0.56$.  Using
$A_V=R_VE(B-V)$ with $R_V=3.1$ (the mean value for the LMC, Cardelli,
Clayton, \& Mathis 1989), the $V$-band extinction would be in the
range of $1.15 \le A_V \le 1.74$.

These values of $A_V \le 1.74$ mag are at odds with the extinction
derived from two other techniques.  In the first case, from our photometry
we find a
$V-K$ color of $V-K=1.17 \pm 0.05$.  
O-stars, including dwarfs, giants, and supergiants of O-subtypes
6 through 9, have
$V-K$ colors in a relatively narrow range of
$-0.87\le V-K \le -0.83$
(Martins \& Plez 2006).  
Assuming $A_K=0.11A_V$ (Cardelli, Clayton, \&
Mathis 1989), we need $A_V=2.28\pm 0.06$ to match the observed $V-K$
color.    
In the second case, the $V$-band extinction can be inferred
from the hydrogen column densities derived from X-ray spectra.  Table
\ref{xray} gives seven determinations of the column density $N_H$ and
the inferred $V$-band extinction, assuming $A_V=N_H/(1.79\pm
0.03)\times 10^{21}$, where $N_H$ is given in units of cm$^{-2}$
(Predehl \& Schmitt 1995).  The values of the extinction cover a
considerable range: $2.57 \pm 0.12 \le A_V \le 4.53\pm 0.29$.  We
believe the first measurement given in the table ($A_V=2.57\pm 0.12$
from Cui et al.\ 2002) is the most reliable given the model used (the
mult-temperature disk blackbody model) and the data (high resolution
grating spectra
from {\em Chandra} HETG).  This measurement differs by the extinction
derived from the $V-K$ color by only $2.5\sigma$.
Furthermore, we note that the X-ray
column density may systematically exceed the value inferred
from optical reddening (i.e.\ dust)
considerations, if even a fraction of the dense stellar wind from
the O-star is not completely ionized.

Adopting $A_V=2.28$ mag, the range of color excess values quoted above
($0.37 \le E(B-V) \le 0.56$) gives $4.07 \le R_V \le 6.16$.  Our
measured value of color excess of $E(B-V)=0.44$ gives $R_V=5.18$.  The
nominal average value in the LMC is $R_V=3.1$, although certain
lines-of-sight can have $R_V$ values that differ significantly from
the average (Cardelli, Clayton, \& Mathis 1989).

\begin{figure}
\epsscale{0.85}
\plotone{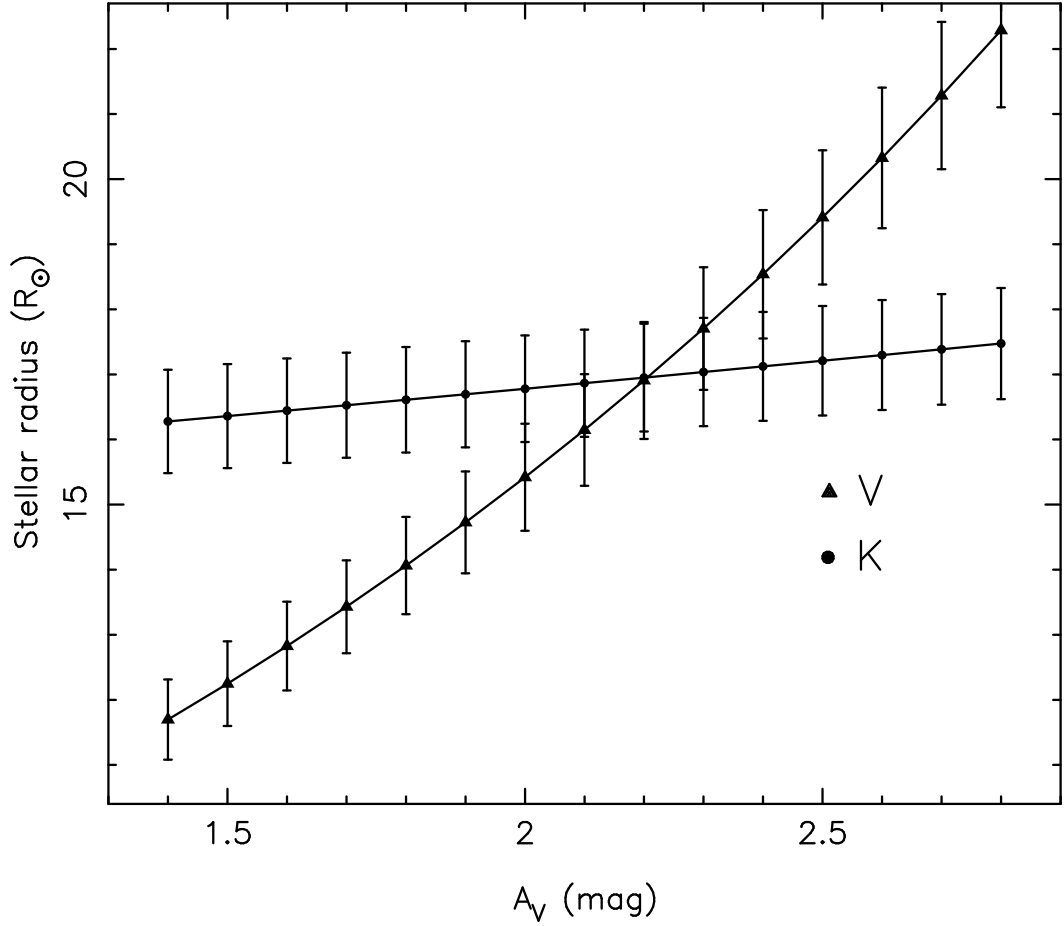}
\caption{The radius of the companion as a function of the extinction
$A_V$ for the $V$-band (triangles) and the $K$-band (circles).  The
radius is the same for both bands when the extinction is 
$\approx A_V=2.2$
mag.}
\label{radfig}
\end{figure}

To compute the radius of the star, we assume a distance modulus to the
LMC of $18.41\pm 0.10$ (see Orosz et al.\ 2007).  Although the
uncertainty of the distance modulus to the center of the LMC itself is
somewhat smaller than 0.1 mag, we adopt an uncertainty of 0.1 mag to
account for the unknown relative position of LMC X-1 with respect to the
LMC center.  We use bolometric corrections for the $V$ and $K$ bands
derived from the OSTAR2002 models with the LMC mean metalicity (Lanz \&
Hubeny 2003).\footnote{The OSTAR2002 bolometric corrections are in good
agreement with the bolometric corrections given in Martins \& Plez
(2006).  We prefer the former since they are tabulated in terms of the
effective temperature and gravity, although the bolometric corrections
are insensitive to the gravity.}  
We use apparent magnitudes of $V=14.6\pm 0.05$ and $K=13.43\pm 0.03$
(where we have increased the uncertainties to be conservative), and
assume the star is the only source of optical light in the system
(see the discussion in the Appendix regarding the likely faintness of
the accretion disk).  
We use an
effective temperature of $T_2=33,200\pm 500$~K, and set the gravity to
$\log g=3.50\pm 0.02$, as determined from the dynamical model described
below.  We computed the stellar radius using the $V$-band magnitudes and
$K$-band magnitudes using a wide range of values of $A_V$, and the
results are shown in Figure \ref{radfig}.  The lines cross near
$A_V\approx 2.2$ mag (where $R_2\approx17\,R_{\odot}$), which is not at
all surprising since we derived $A_V=2.28 \pm 0.06$ mag from the $V-K$
color above.
Note that the radius derived from the $K$ band is relatively insensitive
to
the assumed value of $A_V$, owing to the fact that $A_K=0.11 A_V$.  The
radii found using the $K$-band are in the range $15.5\lesssim R_2
\lesssim 18.3\,R_{\odot}$ (this includes the $1\sigma$ errors) for
extinction values in the generous range of $1.4\le A_V \le 2.8$ mag.
For a given value of $A_V$, the uncertainty in the radius derived from
the $V$ band is on the order of $1.0\,R_{\odot}$ when the generous
uncertainties in the apparent magnitudes and the large range in
effective temperatures are used.  For the $K$ band, the corresponding
uncertainty is on the order of $0.8\,R_{\odot}$.  For the purposes of
the dynamical model below we adopt $R_2=17.0\pm 0.8\,R_{\odot}$, which
is the value we find using the $K$-band and $A_V=2.28 \pm 0.06$.  The
luminosity of the star is then $\log L/L_{\odot}=5.50 \pm 0.05$.

\begin{figure}
\includegraphics[scale=.7,angle=-90]{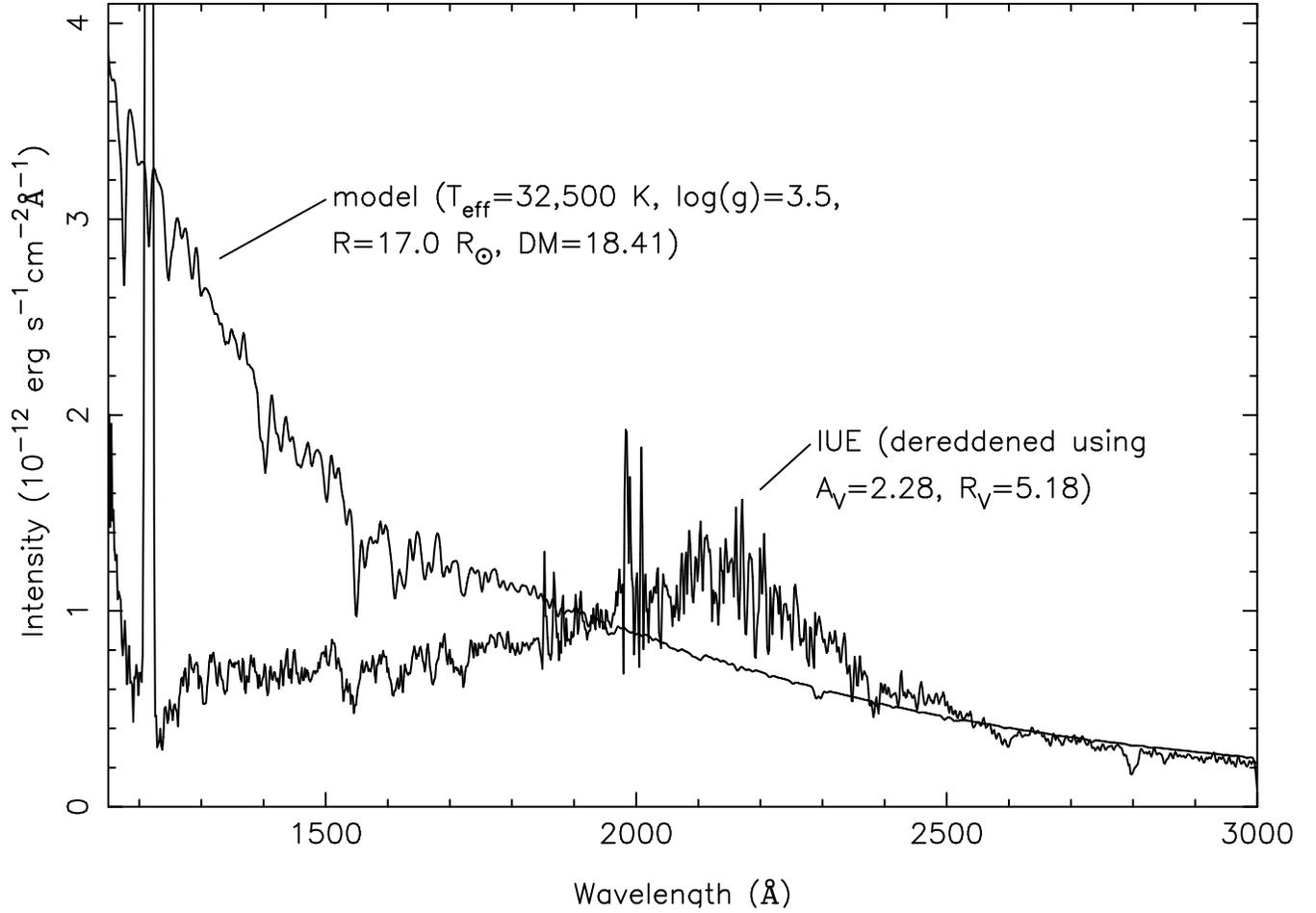}
\caption{The average IUE spectrum of LMC X-1, dereddened using
$A_V=2.27$ and $R_V=5.16$, is shown with a model spectrum from the
OSTAR2002 grid having $T_{\rm eff}=32,500$~K and $\log g=3.5$, scaled
using a distance modulus of 18.41 and a stellar radius of
$17.08\,R_{\odot}$.}
\label{plotiue}
\end{figure}

As noted above, the extinction derived from the IUE spectra is
unusually low.  In order to see if the IUE spectra could be explained
by a different extinction law, we obtained the IUE spectra of LMC X-1
from the MAST archive.  The spectra have been reprocessed using the
latest calibrations, and we combined all of the observations to
increase the signal-to-noise.  The mean IUE spectrum was dereddened
using $A_V=2.28$ and $R_V=5.18$.  The results are shown in Figure
\ref{plotiue}.  We also show a model spectrum from the OSTAR2002 grid
with $T_{\rm eff}=32,500$~K and $\log g=3.5$, scaled using a distance
modulus of 18.41 and a stellar radius of $17.0 \,R_{\odot}$.  The
agreement between the data and the model is good redward of the
2200~\AA\ bump.  However, the 2200~\AA\ bump itself seems
over-corrected, and the slope of the dereddened spectrum blueward of
the bump is much too flat.  However, the following caveats should be
noted regarding the IUE data.  First, the signal-to-noise of the
spectra is not terribly high (on the order of 10 per pixel or less).
Second, LMC X-1 was observed in the IUE large aperture ($10''$), and
the source was not centered in the aperture in order to exclude light
from the much brighter star R148.  Finally, Pakull \& Angebault (1986)
showed that LMC X-1 is in a compact He III nebula, which is in turn
embedded in a larger H II region known as N159 (Bianchi \& Pakull
1985).  Considering the very large aperture of the IUE spectra, the UV
spectrum of LMC X-1 might be contaminated by scattered light from nearby
bright stars.  It would be worthwhile to obtain additional UV spectra
of LMC X-1 to further investigate the UV extinction.

\subsection{System Parameters}\label{parmsec}

\subsubsection{Simultaneous Fits to Light and Velocity Curves}\label{ELC}

We have several ``observables'' of the LMC X-1 binary system (e.g.\
the radial velocity curve of the secondary, its $B$ and $V$ light
curves, etc.), and we seek the physical model whose observed
properties best match the observed data.  While the ``forward''
problem of computing the observable properties of a binary system is
relatively straightforward, the ``inverse'' problem of deriving a
physical model from observed data is much more challenging.  The ELC
code (Orosz \& Hauschildt 2000) is a comprehensive code for computing
the forward problem using a model based on Roche geometry.
ELC can also solve the inverse problem using its associated optimizing codes
based on various numerical techniques
(e.g.\ a ``grid search'', an ``amoeba'', a Levenberg-Marquardt scheme,
a genetic code, and a Monte Carlo Markov Chain scheme).  

The ELC model as applied to LMC X-1 has several parameters, many of
which can be set to reasonable values based on various observed
properties of the system.  The important parameters are related to
either the geometrical properties of the system or to the radiative
properties of the system.

The free parameters which control the basic system geometry include
the orbital period $P$, the orbital separation $a$ (which then gives
the total mass via Kepler's third law), the ratio of the masses
$Q=M/M_2$, the
inclination $i$, and the
Roche lobe filling factor $f_2$ of the
secondary star.  
We find that the search of
parameter space is made more efficient if the $K$-velocity of the
secondary star $K_2$ and the mass of the secondary star $M_2$ are used
to set the system scale instead of $Q$ and $a$.  When modeling actual
data, one must also specify the phase zero point $T_0$, which is
either the time of the inferior conjunction of the secondary star (in
the case of a circular orbit) or the time of periastron passage (in
the case of an eccentric orbit).  We initially
assume the secondary star is
rotating synchronously with its orbit, and that the star's rotational
axis is perpendicular to the orbital plane.
We discuss in
\S\ref{systematic} how our results depend on the assumption of
synchronous rotation.

The parameters which control the radiative properties include the
average temperature of the secondary star $T_2$, its gravity darkening
exponent $\beta$, and its bolometric albedo $A$.  Following standard
practice, we set $\beta=0.25$ and $A=1$, which are values appropriate
for stars with a radiative envelope.  ELC uses specific intensities
derived from model atmosphere calculations, so no parameterized limb
darkening law is needed.  
In modeling the LMC
X-1 data, we use model atmospheres computed for the LMC metalicity
from the OSTAR2002 grid (Lanz \& Hubeny 2003), supplemented by models
from the BSTAR2006 grid at lower temperatures (Lanz \& Hubeny 2007).

ELC can include light from a flared accretion disk, and geometrical
effects due to the disk (e.g.\ eclipses of the secondary star).  If the
disk contributes a significant amount of the light from the binary the
observed amplitude of the ellipsoidal light curve will decrease.  Thus
it is critical to understand how luminous the accretion disk is relative
to the star.  We show in the appendix that the accretion disk
contributes at most $\approx 10\%$ of the light in the optical and
most likely contributes less than
a few percent of the light.  We
therefore omit the disk in the models.  We discuss in \S\ref{systematic}
how our results depend on the assumption of no disk light.

The effects of X-ray heating are computed 
using a scheme adapted from Wilson (1990).  The source of the
X-rays is assumed to be a thin disk with a radius that is 
very small compared to the orbital separation.
The extent to
which X-ray heating changes the light (and velocity) curves depends
primarily on how large the X-ray luminosity is compared to the
bolometric luminosity of the secondary star.  From our ongoing work on
modeling 53 {\it RXTE} PCA X-ray spectra of LMC X-1, we find that the
isotropic bolometric X-ray luminosity of LMC X-1 is quite steady over
a time span of several years and is $\approx 2.3 \times
10^{38}$\ergsec.  For comparison, the bolometric luminosity of the
star is $L\approx 1\times 10^{39}$ erg s$^{-1}$.  
Because the star intercepts only a modest fraction of the total 
X-ray flux, we thus
expect that
the effects of X-ray heating on the light and velocity curves will be
minor.  
We show in \S\ref{systematic} that our main results are
insensitive to reasonable changes in $L_x$, and that systematic errors
in the inclination and the derived masses due to improper treatment of
X-ray heating are very small.
In the following analysis we adopt $L_x=2.3\times 10^{38}$ erg s$^{-1}$.

To summarize, we have seven free parameters: $i$, $K_2$, $M_2$, $P$,
$T_0$, $T_2$, and $f_2$.  The data we model include 58
$B$-band and 57 $V$-band measurements from SMARTS, and 37 radial
velocity measurements.  The $J$-band light curve from SMARTS proved to
be too noisy to model.  The uncertainties on the individual
observations were scaled to give $\chi^2\approx N$ for each set
separately.  The mean of the error bars after the scaling are 0.012
mag for $B$, 0.015 mag for $V$, and 2.62 km s$^{-1}$ for the radial
velocities.  Finally, we have three additional observational
constraints that we include in the model ``fitness'' (see Orosz et
al.\ 2002): the projected rotational velocity of the star ($V_{\rm
rot}\sin i=129.9\pm 2.2$ km s$^{-1}$), the radius of the secondary star
($R_2=17.0 \pm 0.8\,R_{\odot}$), and the fact that the X-ray source
is not eclipsed.

\begin{figure}
\includegraphics[scale=.7,angle=-90]{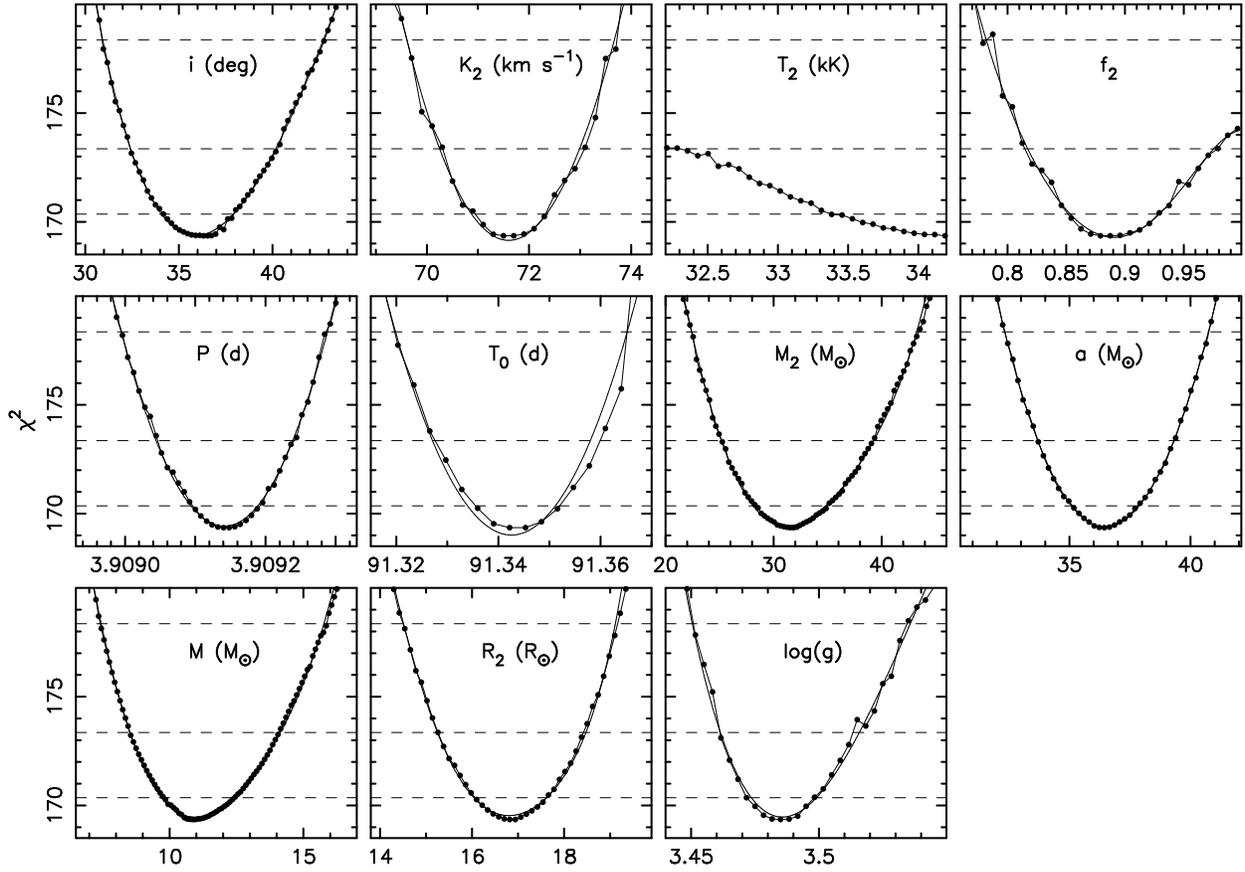}
\caption{Curves of $\chi^2$ vs.\ various fitting and derived
parameters of interest for the circular orbit model.  The horizontal
dashed lines denote the 1, 2, and $3\sigma$ confidence limits.}
\label{plotfitted}
\end{figure}

Parameter space was thoroughly explored by running ELC's genetic
optimizer several times with the order of the parameters changed (this
gives different initial populations while leaving the volume of the
explored parameter space unchanged).  The value of $T_2$ was confined to
the $2\sigma$ range $32,200\le T_2\le 34,200$~K as determined from the
spectra, while the other parameters were given generous ranges.  The
final values of the parameters were refined using the grid search
optimizer.  The uncertainties on the fitted and derived parameters were
found using the technique described in Orosz et al.\ (2002).  
Table \ref{parm} gives the fitted parameters, some derived parameters,
and their uncertainties for our adopted circular orbit model and for an
eccentric model discussed in more detail below.  
Figure \ref{plotfitted} shows curves of $\chi^2$ vs.\
various fitting and derived parameters of interest for the circular
orbit model, from which one may read off the best-fitting value of a
given parameter and
the $1\sigma$, $2\sigma$, and $3\sigma$ confidence ranges.  Each
parameter is well constrained by the ellipsoidal model, with the
exception of the mean temperature of the secondary star $T_2$.  
The fact
that $T_2$ is not constrained by the ellipsoidal model is not
surprising in this case (e.g.\ a single optical star in a non-eclipsing
system) and is not a cause for concern since the optical spectra
constrain the temperature quite nicely.  We discuss in
\S\ref{systematic} how our results depend on the adopted value of $T_2$
for a wide range of values.
Figure \ref{lcfig1} shows the folded light curves with the
corresponding model curves,
while the folded radial velocity curve and the
corresponding model curves are shown in Figure \ref{rvfig1}.

\subsubsection{Possible Systematic Errors in the Light Curve 
Models}\label{systematic}

Our adopted parameters depend, among other things, 
an assumption that the orbit is circular and the star rotates
synchronously with the orbit, and
on two measured
quantities that are not directly constrained (or only weakly
constrained) by the light or velocity curves: the mean temperature of
the secondary star $T_2$ and its radius $R_2$.  The radius of the
star, when used as an observed quantity, helps constrain the scale of
the binary and hence the component masses.  The temperature is needed
to find the radius independently.  
We did numerical experiments to
see how our results change when nonsynchronous rotation
and eccentric orbits are allowed, when
the input temperature of the star is
changed, and when the radius (used as an extra constraint) is changed.
We also explored how our results depend on our assumption that 
there is no disk light contamination in the optical and how our results
change with different assumed values of the X-ray luminosity.
But first, we briefly discuss the slight difference between the $K$-velocity
derived from simple sinusoid fits to the radial velocities and
the $K$-velocity found from the joint radial velocity and light curve fits.

\paragraph{Systematic Errors in $K_2$.}

The estimated $K$-velocity for LMC X-1 depends slightly on which lines
are used and how the spectra are normalized (\S\ref{radvel} and
Fig.\
\ref{Kfig}).
A four-parameter sinusoid fit to the thirty seven
radial velocities measured using all of the suitable lines 
yields an orbital period of $P=3.909139\pm 0.000054$ d,
and $K_2=69.79\pm 0.65$ km s$^{-1}$, with  $\chi^2=36.50$.
This $K$-velocity is
slightly smaller than the $K$-velocity found from the
detailed modeling of the light and velocity
curves ($K_2=71.61 \pm 0.67$ km s$^{-1}$).
[For comparison Hutchings et al.\ (1983) derived $K_2=65\pm 7$
km s$^{-1}$ for a period of $P=3.909$ d.]
There are two reasons for the difference.  The main
reason is that the orbital phases of the radial velocities are tightly
constrained when they are modeled together with the light curves.  In
this case, the time of the inferior conjunction of the secondary star
found from a simple sinusoid fit to the velocities is slightly
different from the same time found from the combined light and
velocity curve modeling (see below).  
The second reason for the slight difference
in the $K$-velocities is that ELC computes corrections to the radial
velocities due to the distorted nature of the star, so the model curve
is not precisely a sinusoid, even for a circular orbit.

Fitting the radial velocities along with the light curves provides a
physical scale for the model.  If one knew the $K$-velocity ahead of
time, then one in principle could fit the light curves by themselves.
However, since the $K$-velocity is not known exactly, it is better to
use an additional $\chi^2$ constraint for $K_2$ rather than a fixed input
value.  We ran fits to the light curves by themselves using several
values of $K_2$ as an extra constraint, assuming an uncertainty of 0.65
km s$^{-1}$.  The period was fixed at the value found from the
four-parameter sinusoid fit, leaving six free parameters.  The results
are shown in Table \ref{fixK}.  Not surprisingly, the masses scale as
$K_2$.  Since the computation of the light curves is not completely
independent of the scale of the system (the model atmosphere specific
intensities are tabulated in terms of the temperature and gravity in cgs
units), there is a little bit of variation in the derived values of $i$,
$T_0$, and $\chi^2$.
The most important result contained in Table \ref{fixK} is the fact that
the
derived values of the time of inferior conjunction of the secondary
determined from the light curves is $\approx 2\sigma$ different than
that derived from the four-parameter sinusoid fit to the radial
velocities.  Since the photometrically determined time of conjunction
should agree with the spectroscopically determined one, it is clearly
better to fit the radial velocities together with the light curves.  The
$K$-velocity derived from the simultaneous fit will be closer to the
actual value, but at the expense of a larger $\chi^2$ value contribution
from the radial velocities.

\paragraph{Possible Nonzero Eccentricity and Nonsynchronous Rotation.}

Although the timescales for a close binary to evolve to a circular orbit
and synchronous rotation are relatively short, we should nevertheless
entertain the possibility that the orbit of LMC X-1 could be slightly
eccentric and/or the rotation could be nonsynchronous because the age of
the massive secondary star in LMC X-1 might be less than the timescales
for synchronization or circularization.
Also, some examples of high mass
X-ray binaries with eccentric orbits are known and include Vela X-1
(Rappaport, Joss, \& McClintock 1976; Bildsten et al.\ 1997) and M33 X-7
(Orosz et al.\ 2007).  Having an eccentric orbit adds the eccentricity
$e$ and the argument of periastron $\omega$ as free parameters (see Avni
1976 and Wilson 1979 for a discussion of generalized Roche potentials
for eccentric orbits).  As before, parameter space was thoroughly
explored by using the genetic optimizer, and the results are shown in
Table \ref{parm}.

Avni (1976) also discussed the generalization of the Roche potential for
nonsynchronous rotation.  The figure of the star depends somewhat on how
fast it is rotating, and a convenient parameterization of the rotation
rate is $\Omega$, which is the ratio of the rotation frequency of the
star to the orbital frequency.  
For a circular orbit and synchronous rotation, we have $\Omega=1$.
For an eccentric orbit, the idea of synchronous rotation is somewhat
harder to define because the angular orbital speed of the star changes
over its orbit.
One can compute $\Omega$ so that the rotation frequency
of the star matches its orbital frequency at periastron (Hut 1981).
We computed models for a wide range of $\Omega$ for both circular
and eccentric orbits, and the results are given in Table 
\ref{nonsync}.

The eccentricity is formally nonzero at the $\approx 3\sigma$ level
since the total $\chi^2$ of the fit decreased by about 14 with
the addition of two more free parameters.
However, there are several reasons for believing that the nonzero
eccentricity is spurious: 
(i) The sampling of the radial velocities is such that a best-fit sinusoid
is shifted in phase relative to the light curves.  In an eccentric
orbit, the phases of the minimum or the maximum velocity are shifted
relative to a sine curve (see Fig.\ \ref{rvfig1}), so that in the
combined light+velocity curve modeling the total $\chi^2$ can be lower.
Indeed,  
the $\chi^2$ value of the fit to the radial velocities for the eccentric
model (Table \ref{parm}) in the combined light+velocity curve analysis
is
comparable
to the $\chi^2$ of the
four-parameter best-fit sinusoid fit discussed above.  
In other words, 
an eccentric orbit fitted to the radial velocities 
{\em alone} is not significantly better
than a simple sine curve fit.
(ii) The argument of periastron $\omega$ is
consistent with 270$^{\circ}$.  This gives one pause because tidal
distortions of the secondary can give rise to distorted line profiles,
which in turn give rise to systematic errors in the measured radial
velocities.  These velocity errors can result in fits with spurious
eccentricities and with the argument of periastron at a quadrature phase
(e.g.\ Wilson \& Sofia 1976; Eaton 2008).
ELC does compute corrections to the radial velocities in
the manner of Wilson \& Sofia (1976), but these corrections may not be
completely accurate.  (iii) When nonsynchronous models are considered
(Table \ref{nonsync}) we find that the eccentricity is correlated with
the
parameter $\Omega$.  As the value of $\Omega$ gets smaller (i.e.\ as the
star rotates slower and slower), the eccentricity and the inclination
get larger.  In fact, values of $\Omega$ smaller than about 0.65 are not
possible since the best-fitting inclination would produce an X-ray
eclipse, which is not observed at a high level of confidence.  
Hence, we would have for the best-fitting model in Table 5 a star
rotating at $\approx 65\%$ of
the synchronous value in a moderately eccentric orbit where the X-ray
source passes just below the limb of the star as seen in the plane of
the sky.  
If we ignored the fact that there is no X-ray eclipse, our best-fitting
model would have $i=90^{\circ}$.
(iv) The differences between the model light curves for the
eccentric orbit model and the circular orbit model are quite subtle.
(v) Detailed model
computations specific to LMC X-1 give timescales for synchronization and
circularization that are both a few orders of magnitude shorter than the
lifetime of the secondary star for most values of the initial rotation
rate (Valsecchi, Willems, \& Kalogera, private communication 2008).

Given these arguments which cast doubt on the reality of a nonzero
eccentricity, we
adopt the parameters derived from the circular orbit model.  To be
conservative with the uncertainties, 
the
small differences between the parameters derived from the circular
orbit model and the eccentric orbit model are taken to be a measure of
the systematic errors.  These systematic errors are added in
quadrature to the uncertainties found for the circular orbit model to
produce the final adopted uncertainties (fourth column in Table
\ref{parm}).

Regarding the  models for nonsynchronous rotation, some trends
are evident in Table \ref{nonsync}.  First, as $\Omega$ goes from high
values to lower values, the mass of the secondary star and the orbital
increase, and the mass of the black hole decreases (i.e.\ the mass ratio
$Q$ decreases).
The reason for the
mass ratio change is easy to understand.  One can show that the
rotational velocity of a star that fills its Roche lobe and is in
synchronous rotation depends on the mass ratio (Wade \& Horne 1988). The
same holds true for a star that underfills its Roche lobe, although the
function relating the velocity to the mass ratio becomes more
complicated.  When the rotation is nonsynchronous, the relation between
the velocity and the mass ratio changes further.  The reason for the
change in the inclination is also easy to understand.  
As the value of $\Omega$ decreases (i.e.\ as the star rotates more
slowly), the star becomes less distorted.  As a result,
higher inclinations are needed to get roughly the same amplitudes
for the light curves.

Finally, as the value of $\Omega$ goes down, the value of $\chi^2$ for
the fit goes down as well.  However, the change in $\chi^2$ over the
entire range of $\Omega$ for the circular orbit is modest
($\Delta\chi^2\approx 3.5$), whereas for the eccentric orbit the change
is much larger ($\Delta\chi^2\approx 14$).  
We have already discussed why we believe that the non-zero eccentricity
may be spurious.
For the circular
orbit model, we have no strong observational constraint on the value of
$\Omega$.  Based on other considerations, we give two reasons why we
believe $\Omega=1$.  
First, our formal best solution corresponds to a near-grazing X-ray
eclipse, which although
possible, seems unlikely.  
Second, as discussed above, the timescales
for both synchronization and circularization are much shorter than the
age of the secondary star for most initial conditions.  We conclude that
$\Omega$ is most likely to have a value very close to unity.

\paragraph{Fixed Temperature.}

In our next experiment, we imagine that the temperature of the
secondary is known exactly.  In this case, the bolometric correction
is known exactly (to the extent that one believes the model atmosphere
computations).  Once the bolometric correction is known, the radius of
the star can be found using the apparent $K$ magnitude, the distance,
and the extinction $A_V$.  We ran eight simulations where the
temperature was fixed at values between 30,000~K and 37,000~K in steps
of 1000~K.  We assumed a distance modulus of $18.41\pm 0.10$ mag and
an extinction of $A_V=2.28\pm 0.06$ mag.  
Table \ref{fixT} gives the derived radii and luminosities for
these eight cases.  For each case, we fit the data using the genetic
code, assuming the orbit is circular.  Table \ref{fixT} also
gives the resulting values of the inclination $i$, secondary star mass
$M_2$, the black hole mass $M$, and the $\chi^2$ of the fit.  A few
trends are obvious from Table \ref{fixT}.  First, as the temperature
increases, the derived radius is roughly constant
until $T=34,000$, after which it decreases.   
This is due in part to the way the bolometric corrections
change with temperature.  The luminosity also increases with increasing
temperature.
Second, as the temperature increases, the 
$\chi^2$ of the fit generally goes down.  
On the other hand, the derived component masses do not change that
much over the range of temperatures.  The extreme values of the black
hole mass are $8.79\pm 0.75\,M_{\odot}$ and 
$11.21\pm 1.35\,M_{\odot}$.

\paragraph{Fixed Radius.}
Next, we can imagine that the radius of the
secondary star is known exactly.  In this case, we do not need to know
either the distance to the source or the extinction.  Using a
temperature of $T_2=33,200\pm 500$~K, we can compute
the luminosity of the star and its uncertainty.  
We ran eight simulations. In each we fit the data using the genetic
code (again using a circular orbit) with the secondary star radius
fixed at one of the eight equally-spaced values between
$15.0\,R_{\odot}$ and $19.0\,R_{\odot}$.
The results of this exercise (see  Table \ref{fixR}) 
can be used to judge what would happen
to our results if our adopted distance to the LMC is in error, or if
our adopted value of the extinction $A_V$ were in error.  For example,
if the LMC were actually closer than what we assume, the radius that
we would derive would be smaller (with all other things being equal).
As
before, a few trends are evident when inspecting Table \ref{fixR}.
First, it is obvious that the luminosity of the star will go up as its
radius goes up.  As the radius increases, the scale of the binary
increases leading to larger masses.  As the star's radius goes
up, it fills more of its Roche lobe, resulting in lower inclinations.
In fact, 
good solutions with 
radii larger than $\approx 19.0\,M_{\odot}$ are not possible since the
star would then
overfill its Roche lobe.  Unlike the case when the temperature
was fixed at various values, the $\chi^2$ of the fit changes very
little when the radius is fixed at different values, except when
$R_2\gtrsim 18\,R_{\odot}$.

\paragraph{Variations in X-ray Heating.}
As noted above, ELC assumes the source of X-rays is a thin disk
with a small diameter compared to the orbital separation.
Our value of $L_x$ ($\approx 2.3\times 10^{38}$ erg
s$^{-1}$,
which includes the geometrical factor to account for the inclination
angle of the disk as seen from Earth) may need two correction factors  
when
used as an input to the simple way in which ELC computes the effects
of X-ray heating.  First, if there is extinction local to the source
(for example due to the outer edge of the accretion disk),
some of the X-rays that would have reached the secondary star could be
absorbed.  If a flared accretion disk is present in the model, ELC
can test which grid elements on the star are shielded from the X-ray
source by the disk rim.  However, it is hard to know what parameters
to choose for the disk without more observational constraints.  
In any case, the first correction factor would be less than unity.
The
second correction factor comes from the fact that the X-ray source 
may not be a perfectly flat and thin disk
that is in the orbital plane (i.e.\ the disk may be warped or there
may be some kind of spherical corona that emits some of the X-rays).
When there is a perfectly thin and flat disk, 
an observer on the star would see the X-ray
emitting thin disk at an angle, and the intensity of X-rays observed
at that point would be lower than it would be for the simple point source
case owing to foreshortening of the disk (the foreshortening of the
surface element itself is accounted for in the code).  
A warp in the disk or the presence of a quasi-spherical corona could
increase the effective emitting area, leading to more X-rays hitting the
secondary star.  The second correction factor would be greater than
unity.

To see how much the derived parameters depend on our
adopted value of $L_x$, we ran nine numerical experiments with $\log
L_x$ fixed at values between 38.0 and 39.0 in steps of 0.25 dex.  We
also set $L_x=0$ (e.g.\ the case for no X-ray heating).  We assume a
circular orbit, and that the free parameters and the ranges thereof were
the same as in our main work.  Table \ref{fixX} gives the results.  
As
the value of $L_x$ goes up, the $\chi^2$ of the fit remains virtually
the same.
In addition, there is little variation in the inclination and in
the derived masses, and we conclude that our overall results are
rather insensitive to our treatment of the X-ray heating.

\paragraph{Variations in the Disk Contribution.}
We show in the Appendix that the accretion disk contributes a few percent
or less of the flux in the optical bands, and at most $\approx 10\%$
when extreme ranges of parameters are considered.  In the analysis above
we assumed there was zero disk contribution.
We ran models which included an accretion disk to see how our
results depend on this assumption of zero disk contribution
in the optical.  In the ELC code one must specify the inner and outer 
disk radii, the disk opening angle, the temperature at the inner edge,
and the exponent on the power-law temperature profile.
Since LMC X-1 is not eclipsing, there is little
to constrain the parameters of the accretion disk.  Consequently we
fixed the inner radius, the opening angle, and the exponent
on the power-law temperature profile.  The genetic code was run
four times, and
models where the disk fraction was within small threshold of a specified
value were selected by means of an additional term in the
total $\chi^2$.   The various runs had the disk fraction in $V$ set at
0.05, 0.10, 0.15, and 0.20.
If there is optical light from the accretion disk, then the O-star
by itself would be less luminous since the apparent magnitude of the
disk+star combination is fixed.  As a result its computed radius would
be smaller for
a fixed temperature. 
Therefore   
in each of the four cases
the assumed radius of the O-star was adjusted downward
accordingly.
The results are shown in Table \ref{diskfrac}.
In general, as the disk fraction increases, the inclination increases
slightly
and the masses go down slightly.   For a disk fraction of 10\%, the mass
of the black hole decreases by $\approx 1\sigma$.  We therefore conclude
that our 
results are relatively insensitive to our assumption of no disk
contribution in the optical.

\section{Discussion}\label{discuss}

\subsection{The Mass of the Black Hole}

The precision of our measurement of the black hole mass ($10.91\pm
1.55\,M_{\odot}$) represents a great improvement over the pioneering
work of Hutchings et al.\ (1987) who give a mass ``near
$6\,M_{\odot}$'' for the black hole.  
Prior to this work, relatively
little progress was made on the dynamics of LMC X-1; for example,
in their recent compilation Charles \& Coe
(2006) give a range of $8-20\,M_{\odot}$ for the mass of the black
hole in LMC X-1.  
Most of the binaries that contain dynamically confirmed black holes
have low mass secondary stars
(where $M_2\lesssim 2.5\,M_{\odot}$), and the sample size of
black hole binaries with high mass secondaries 
(where $M_2\gtrsim 10\,M_{\odot}$),
is rather limited at
the moment.
There are three systems with high mass secondaries where dynamical
studies unequivocably show that the compact object must be a black
hole: LMC X-1, M33 X-7, and Cyg X-1.
The component masses of M33 X-7 are well constrained ($M=15.65\pm
1.45\,M_{\odot}$ and $M_2=70.0\pm 6.9\,M_{\odot}$, Orosz et al.\
2007), whereas there is some disagreement on the parameters of Cyg X-1
(Herrero et al.\ 1995 give masses of $\approx 10\,M_{\odot}$ and
$18\,M_{\odot}$ for the black hole and secondary star, respectively
while Charles \& Coe give $>4.8\,M_{\odot}$ for the black hole mass).
There are two other eclipsing high mass X-ray binaries to mention
in this context.
4U 1700-377 has a compact object with a mass of
around $2.4\,M_{\odot}$, but this object may be a massive neutron
star (Clark et al.\ 2002).  IC 10 X-1
has a large optical mass function that was measured using the He II
$\lambda4686$ emission line ($7.64\pm 1.26\,M_{\odot}$,
Silverman \& Filippenko 2008; Prestwich et al.\ 2007).  Assuming the
He II line properly tracks the motion of
the Wold-Rayet (WR) secondary star, and assuming a mass for
the WR star based on evolutionary models, the compact object in
IC 10 X-1 has a mass of $\approx 30\,M_{\odot}$ or more.
Fortunately, the future for studies of black hole binaries with high
mass secondaries appears promising.
Deep X-ray surveys of nearby galaxies are finding more and more
high mass X-ray binaries that can be followed up in ground-based
studies with the new generation of large telescopes and high-performance
focal-plane instrumentation.

\subsection{Evolutionary Status of the Secondary Star}\label{evol}

\begin{figure}
\epsscale{0.85}
\plotone{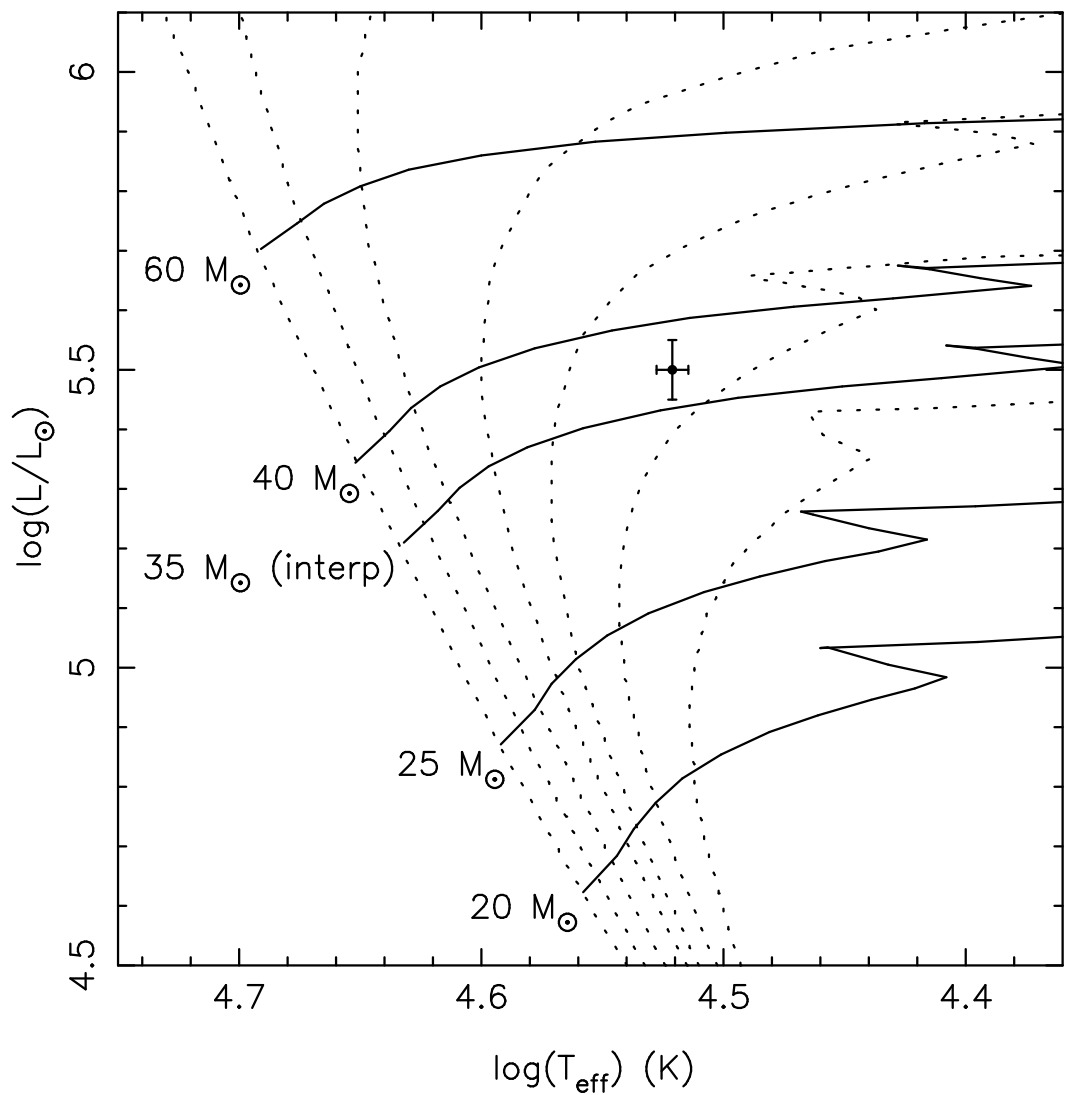}
\caption{The solid lines show evolutionary tracks for an LMC
metalicity on a temperature-luminosity diagram for stars with ZAMS
masses of $60\,M_{\odot}$, $40\,M_{\odot}$, $25\,M_{\odot}$,
$20\,M_{\odot}$ taken from Meynet et al.\ 1994).  The track for a
ZAMS mass of $35\,M_{\odot}$ was crudely interpolated.  The dotted
lines show isochrones (also for an LMC metalicity, Lejeune \& Schaerer
2001) from left to right of 0, 1, 2, 3, 4, 5, and 6 Myr.  The location
of the LMC X-1 secondary is very close to the interpolated track for
the $35\,M_{\odot}$ star and the 5 Myr isochrone.}
\label{plothr}
\end{figure}

Using our measured temperature and luminosity of the secondary star
star ($33,200\pm 500$~K, and $\log L/L_{\odot}=5.50 \pm
0.05$) we can place the star on a temperature-luminosity
diagram (Figure \ref{plothr}).  The position of the star in the
diagram can be compared with theoretical evolutionary tracks for
single stars (Meynet et al.\ 1994) and with theoretical isochrones
(Lejeune \& Schaerer et al.\ 2001).  
Although we recognize that
the evolutionary models for massive stars are uncertain
owing to assumptions regarding stellar winds and rotation, 
for the purposes of this discussion
we will take the evolutionary tracks and the isochrones at face value.
The position of the LMC X-1
secondary is very close to the 5 Myr isochrone.  Meynet et al.\
(1994) give evolutionary tracks for ZAMS masses of 20, 25, 40, and
$60\, M_{\odot}$.  To get a crude idea of where the evolutionary track
of a star with an initial mass of $35\,M_{\odot}$ would be, we plotted
the initial luminosity vs.\ the initial mass and the initial
temperature vs.\ the initial mass for the tabulated models and used
quadratic interpolation to determine the initial temperature and
luminosity of a star with a ZAMS mass of $35\,M_{\odot}$.  The
evolutionary track for the star with a ZAMS mass of $40\,M_{\odot}$
was then shifted down to the interpolated position of the
$35\,M_{\odot}$ star and plotted.  The interpolated track with a ZAMS
mass of $35\,M_{\odot}$ passes very near the position occupied by the
LMC X-1 secondary.  The star with a ZAMS mass of $40\,M_{\odot}$ has
lost about $2.6\,M_{\odot}$ by the time its temperature has fallen to
$\approx 33,000$~K.  If a star with a ZAMS mass of $35\,M_{\odot}$
behaves in a similar manner, then its mass would be $\approx
32.4\,M_{\odot}$ when its temperature has dropped to 33,000~K, which
is consistent with what we measure ($M_2=31.79\pm 3.48\,M_{\odot}$,
Table \ref{parm}).

As one can see from the increasing space between the isochrones, as
the age increases these massive stars move quickly across the
temperature-luminosity diagram.  Since the secondary star in LMC X-1
nearly fills its Roche lobe (the fraction filled is about 90\%), it
will soon enter into an interesting stage of binary star evolution.
In round numbers, the star's current radius is $17\,R_{\odot}$ and
the radius of its Roche lobe is $19\,R_{\odot}$.  From interpolation
of the evolutionary track, the star with a ZAMS mass of
$40\,M_{\odot}$ has a radius of $17\,R_{\odot}$ at an age of 4.05
Myr and a radius of $19\,R_{\odot}$ at an age of 4.22 Myr.  If the
evolution of a star with a ZAMS mass of $35\,M_{\odot}$ proceeds at a
similar pace, the secondary star in LMC X-1 will encounter its Roche
lobe in only a few hundred thousand years.  Since the secondary star
is substantially more massive than the black hole, the resulting mass
transfer will be rapid and
may be unstable (Podsiadlowski, Rappaport, \& Han 2003).
If the mass transfer is stable and the binary survives the initial phase
of rapid mass transfer, the present-day secondary may become detached again
in a binary with a longer orbital period.  On the other hand, 
if the rapid mass transfer is unstable,
the orbit will rapidly shrink
 and the binary will enter a
``common envelope'' (CE) phase.
The outcome of the  CE phase
depends on how tightly bound the envelope of the star is.  If the
envelope is tightly bound, then the energy liberated as the two stars
move towards each other will not be enough to expel the envelope, and
the two stars will merge.  On the other hand, if the envelope is
loosely bound, then the envelope of the star can be expelled before
the cores merge, leaving behind a tight binary consisting of the
present-day black hole and the core of the present-day secondary star.
Although detailed computations are required to assess these scenarios, 
it seems that the most
likely outcome of the  CE phase in LMC X-1 would be a merger
since the envelopes of massive stars are tightly bound, and the
density gradient in a main sequence star is not nearly as steep as it
is in a giant (e.g.\ Podsiadlowski et al.\ 2003).  What the merger
product would look like is an open question.

\subsection{The O-star Wind and the ASM X-ray Light Curves}\label{wind}

\begin{figure}
\epsscale{0.85}
\plotone{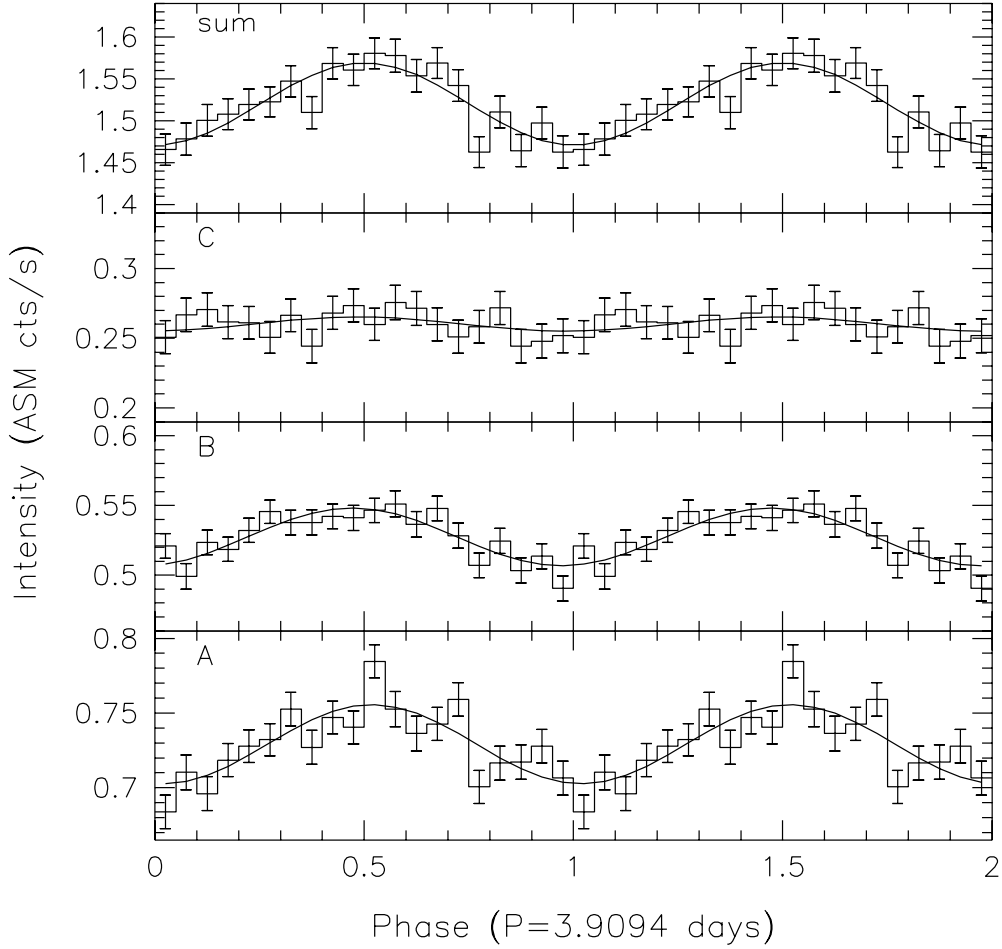}
\caption{The {\it RXTE} ASM light curves of LMC X-1 folded using a
period of 3.9094 days and an epoch of phase zero of MJD 53390.75174.
The four panels from bottom to top show the data from the A (1.5-3
keV), B (3-5 keV), and C (5-12 keV) bands, and the sum of the three
bands (nominally comprising photon energies 1.5-12 keV).  The error
bars indicate $\pm 1 \sigma$ statistical uncertainties.  The smooth
curves are the best-fit functions of the form $f(\phi) = a_0 +
a_1\cos(2\pi\phi) + a_2\sin(2\pi\phi)$ (see Table
\protect\ref{ASMfits} for the best-fitting parameters).}
\label{ASMfold}
\end{figure}

The X-ray minimum/maximum in the ASM X-ray light curves occurs at
inferior/superior conjunction of the secondary star (Fig.\
\ref{ASMfold}), as expected if the modulation is caused by scattering
or absorption in a stellar wind. A simple cosine wave gives a
reasonable fit to the light curves; the respective full amplitudes for
the A, B and C channels are $0.072 \pm 0.010$, $0.077 \pm 0.11$, and
$0.038 \pm 0.029$ (Fig.\ \ref{ASMfold} and Table \ref{ASMfits}). Thus,
the modulation is largely independent of energy, which indicates that
electron scattering is the main source of opacity. We now show that a
simple model of Thomson scattering in a spherically symmetric stellar
wind is not only reasonably consistent with the observed X-ray
modulation but also yields plausible values of the wind mass flux,
accretion rate, and accretion efficiency. For this section we use the
values of component masses and other system parameters in the
right-hand column of Table~\ref{parm}.

We assume that the O star has a radiatively driven wind like those
from other hot stars and that the wind velocity has a radial
dependence that follows a ``beta law'', i.e., $v(r) = v_{\infty}(1 -
b/r)^\beta$ where $b$ is approximately equal to the O-star
photospheric radius $R_2$ \citep[e.g.,][and references
therein]{dess05,kudpuls00}. The terminal velocity $v_{\infty}$ can be
estimated from the escape velocity at the stellar surface and the
photospheric temperature \citep{kudpuls00}. The escape velocity
$v_{\rm esc}$ can be estimated in turn from the mass and radius
results in Table~\ref{parm}.  We obtain $v_{\rm esc} \approx 590 \pm
40$ km s$^{-1}$.  In the calculations below, we take the values of
$v_{\infty}/v_{\rm esc}$ and of $\beta$ to be like those in the wind
model presented for M33 X-7 by Orosz et al.\ (2007), and so we adopt
the values $v_{\infty} = 1400$ km s$^{-1} \sim 2.4 v_{\rm esc}$ and
$\beta = 1$.  This value of $v_{\infty}/v_{\rm esc}$ is appropriate
for an O-star with photospheric temperature $T_2 \approx 33,000$ K
\citep{kudpuls00}.  The wind velocity at the radius corresponding to
the location of the black hole ($r = a = 36.5$ \rsun) is then $v(a)
\sim 740$ km s$^{-1}$.  Adding in quadrature the transverse velocity
of the black hole through the wind, we find that the gas flows by the
black hole at a velocity $V \sim 880$ km s$^{-1}$.  Given the
substantial uncertainties in both the form of the velocity law and the
values of the parameters, these estimates of $v(a)$ and $V$ must each
be uncertain by as much as 50\%.

In any spherically-symmetric wind model, the scattering optical depth
along the line of sight to the X-ray source at superior conjunction
only differs from that at inferior conjunction by the optical depth
along an additional path of length $d = 2a \sin i = 43.3R_{\odot}$ at
superior conjunction. Let $N_{\rm e}$ be the electron column density
along this part of the superior conjunction line of sight. Then there
must be a scattering optical depth along this part of the line of
sight of ${N_{\rm e}}{\sigma_{\rm T}} \approx 0.07$, where
$\sigma_{\rm T}$ is the Thomson cross section. This, in turn, implies
$N_{\rm e} \sim 1.1 \times 10^{23}$ cm$^{-2}$. The mean electron
number density along the path of length d is then $n_{\rm e} \sim 3.5
\times 10^{10}$ e$^-$ cm$^{-3}$. In our adopted wind model, this mean
density is $\sim 33$\% higher than the density at the black hole
orbital radius (and it is also $\sim 33$\% lower than at the point of
highest density along this section of the line of sight). The gas
density at the black hole orbital radius is then $\rho(a) \sim 5.3
\times 10^{-14}$ g cm$^{-3}$. The corresponding mass loss rate in the
O-star wind is $\dot M_{W} = 4\pi{a^2}{\rho(a)}v(a) \sim 5 \times
10^{-6}$ \msun/yr$^{-1}$.

If the wind of the O star in LMC X-1 is only partially ionized like
the winds of ordinary O-stars, then photoelectric absorption should be
manifest via a larger degree of modulation in the lower energy band
ASM light curves. Is it a reasonable assumption in the case of LMC X-1
that iron and the other metals in the wind have been fully stripped of
their K shell electrons? Some support for this idea is provided by the
remarkable photoionized He III nebula that surrounds LMC X-1 (Pakull
\& Angebault 1986). More directly, at every point on the path
described above the ionization parameter (Tarter et al.\ 1969) is
large, $L/nR^2 > 1000$. Furthermore, the spectrum of LMC X-1 is soft
(disk blackbody temperature $kT \approx 0.9$ keV), which makes it an
effective ionizing agent. Under these conditions, studies of similar
high-mass X-ray binaries support our conclusion that the wind is
close to fully ionized in the region in question (McCray et al. 1984;
Vrtilek et al. 2008).

A crude estimate of the accretion rate is given by the rate at which
the matter in the wind enters a cylinder of radius $r_{\rm c}
\approx 2GM/V^2$ centered on the black hole, where $G$ is the
gravitational constant, $M$ is the black-hole mass, and $V$ is the
(above estimated) velocity of the wind relative to the black hole
\citep[see, e.g.,][]{st83}.  The accretion rate is then given by $\dot
M_{\rm B-H} \sim {\pi}r_{\rm cap}^2{\rho(a)}{V}$. Using the value of
$M$ in Table \ref{parm} and the values of $\rho(a)$ and $V$ estimated
above, we obtain $r_{\rm c} \sim 5$ \rsun\ and $\dot M_{\rm B-H}
\sim 3 \times 10^{-8}$ \msun\, yr$^{-1}$ with uncertainties in each
value of roughly a factor of two.

As noted above, we find that the isotropic bolometric X-ray luminosity
of LMC X-1 is quite steady over a time span of several years and is
$\approx 2.3 \times 10^{38}$\,\ergsec. With $M = 10.91$ \msun \ and the
above estimated accretion rate, the implied accretion efficiency is
$\eta \sim 0.1$. This is comparable to the 0.06 efficiency of a
Schwarzschild black hole or the canonical value of 0.1 that is
commonly used, but, considering that this efficiency estimate is
uncertain by at least a factor of two, this may be fortuitous.

\section{Summary}\label{summary}

We present a new dynamical model of the high mass X-ray binary LMC X-1
based on high quality echelle spectra, extensive optical light curves,
and infrared magnitudes and colors.  From our optical data we find an
orbital period of $P=3.90917 \pm 0.00005$ days.  We present a refined
analysis of the All Sky Monitor data from {\em RXTE} and find a period
of $P=3.9094 \pm 0.0008$ days, which is consistent with the optical
period.  A simple model of Thomson scattering in the stellar wind
accounts for the 7\%
modulation seen in the X-ray light curves.  We find
that the $V$-band extinction to the source is much higher than
previously assumed.  The $V-K$ color of the star ($1.17\pm 0.05$)
implies $A_V=2.28\pm 0.06$, whereas several determinations of the
column density from X-ray observations give $A_V>2.57\pm 0.12$.  The
color excess of $E(B-V)=0.44$ together with $A_V=2.27$ gives
$R_V=5.18$, which is much higher than the nominal mean value in the
LMC of $R_V=3.1$.  For the secondary star we measure a radius of
$R_2=17.0 \pm 0.8 \,R_{\odot}$ and a projected rotational velocity
of $V_{\rm rot}\sin i= 129.9 \pm 2.2$ km s$^{-1}$.  Using these measured
properties of the companion star to constrain the dynamical model of
the light and velocity curves, we find 
an inclination of $i=36.38 \pm 1.92^{\circ}$, a
secondary star mass of $M_2=31.79\pm 3.48\,M_{\odot}$, and a black
hole mass of $10.91\pm 1.41\,M_{\odot}$.    The present location
of the secondary star in a temperature-luminosity diagram is
consistent with that of a star with an initial mass of $35\,M_{\odot}$
that is 5 Myr past the zero-age main sequence.  The star nearly fills
its Roche lobe ($\approx 90\%$)
and, owing to the rapid
change in radius with time in its present evolutionary state, it will
encounter its Roche lobe and begin rapid and
possibly unstable mass transfer in only a
few hundred thousand years.

\acknowledgments

This publication makes use of data products from the Two Micron All
Sky Survey, which is a joint project of the University of
Massachusetts and the Infrared Processing and Analysis
Center/California Institute of Technology, funded by the National
Aeronautics and Space Administration and the National Science
Foundation.  CBD, MMD, and MB gratefully acknowledge support from the
National Science Foundation grant NSF-AST 0707627.  The work of JEM
was supported in part by NASA grant NNX08AJ55G.  MB gratefully
acknowledges support from the National Science Foundation through
grant AST 04-53609 for the California State University Undergraduate
Research Experiences (CSUURE) program at San Diego State University.
DS acknowledges a STFC Advanced Fellowship as well as support
through the NASA Guest Observer Program.
We thank Thierry Lanz for providing the OSTAR2002 bolometric
corrections computed for the $K$ band.  We acknowledge helpful
discussions with Bram Boroson and Tim Kallman on photoionized winds.
We thank Phil Massey for helpful discussion on O-stars.

\appendix

\section{Accretion Disk Contamination in the Optical and Infrared Bands}

The optical and infrared bands of interest to us
for the ellipsoidal modeling ($B$, $V$, 
and $J$) are in
the Rayleigh-Jeans part of the spectrum for the emission from both the
secondary star and the disk (see below).  Therefore, the flux we
receive in any particular band is proportional to the product of the
projected area of the radiating surface $A$ and the temperature $T$.

The radius of the secondary star is $17.0\,R_\odot$ and its
photospheric temperature is $33,200\,{\rm K}$.  Therefore, for the
radiation from the star, we have
\begin{equation}
\langle AT\rangle_{\rm sec} = 1.5\times10^{29} ~{\rm cm^2\,K}.
\label{ATsec}
\end{equation}
In the case of the disk, we note that the circularization radius of
the captured wind material is likely to be quite small.  Nevertheless,
the steady state disk will probably have a large outer radius since
the gas needs to get rid of its angular momentum.  An extreme
possibility is that the angular momentum is removed solely by tidal
torques from the companion star acting on the outer regions of the
disk.  From the discussion in Frank, King \& Raine (2002; eq.\ 5.122),
the outer radius of the disk is then approximately equal to 0.9 times
the Roche lobe radius.  Using a standard approximation for the Roche
lobe radius (Frank et al. 2002), we find for the outer radius
\begin{equation}
R_{\rm out} = 0.9\times0.462 \left({M_{\rm BH}\over
M_{\rm BH}+M_{\rm sec}}\right)^{1/3} a =6.6\times10^{11} ~{\rm cm}.
\end{equation}

We further note that most of the disk will be dominated, not by local
viscous dissipation, but by reprocessing of radiation emitted by the
inner region of the disk.  Assuming that $10\xi_{10}\%$ of the
bolometric luminosity of the disk is reprocessed (the exact fraction
is uncertain), the luminosity of the reprocessed radiation from the
disk is $2.3\times10^{37}\xi_{10} ~{\rm erg\,s^{-1}}$.  For a typical
reprocessing geometry, the effective temperature of the reprocessed
radiation will vary as $R^{-1/2}$ (as compared to the steeper
$R^{-3/4}$ variation one expects for viscously generated flux).  Let
us write the disk temperature as
\begin{equation}
T(R) = T_{10}\,(R/10^{10}\,{\rm cm})^{-1/2},
\end{equation}
and let us assume that the reprocessing region of the disk extends
from $R_{\rm in} \sim100R_S = 3.1\times 10^8 ~{\rm cm}$ to $R_{\rm
out}$ (the precise value of $R_{\rm in}$ is not important for what
follows).  Then we have the condition (in cgs units)
\begin{equation}
2.3\times10^{37}\xi_{10} = \int_{3.1\times10^8}^{6.6\times10^{11}}
4\pi R\sigma T_{10}^4 (R/10^{10})^{-2}dR.
\end{equation}
Solving, we obtain
\begin{equation}
T_{10} = 8.1\times10^4\xi_{10}^{1/4} \,{\rm K}.
\end{equation}
The disk temperature thus varies from $4.6\times
10^5\xi_{10}^{1/4}\,$K at $R_{\rm in}$ to $1.0
\times10^4\xi_{10}^{1/4}\,$K at $R_{\rm out}$.  Even the outermost
region of the disk is hot enough that we may treat its optical/IR
emission in the Rayleigh-Jeans limit.

We can now estimate the value of $\langle AT\rangle$ for the disk
emission:
\begin{equation}
\langle AT\rangle_{\rm disk} = \int_{3.1\times10^8}^{6.6\times10^{11}}
2\pi R(\cos i) T_{10}\,(R/10^{10})^{-1/2} dR
=1.5\times10^{28}\xi_{10}^{1/4} ~{\rm cm^2\,K}.
\label{ATdisk}
\end{equation}
Comparing this estimate with equation (\ref{ATsec}), we see that the
disk will contribute about 10\% of the flux in the optical and
infrared.  Note that the uncertain parameter $\xi_{10}$ appears only
with a 1/4 power, so we are not sensitive to its value.  However, the
result is very sensitive to the assumed outer radius of the disk.  If
the disk is smaller than we have assumed, the disk flux will be
reduced substantially, roughly as $R_{\rm out}^{3/2}$.

For a wind-fed system like LMC X-1, the disk will probably get rid of
much of its angular momentum by interacting directly with the incoming
material.  The efficiency of this process is very difficult to
estimate from first principles, so we will use the eclipsing system
M33 X-7 as a guide to what the outer radius 
of the accretion disk
might be.  Orosz et al.\ (2007) determined an outer radius 
of $0.45\pm
0.03$ times the Roche radius for M33 X-7.  If the accretion disk in
LMC X-1 fills 45\% of its Roche lobe then the disk contribution will
be $\langle AT \rangle_{\rm disk} = 5\times 10^{27}\xi_{10}^{1/4}$,
which is about 3.5\% of the optical/infrared flux.  Even this may be
an overestimate.  It is quite conceivable that the disk is much
smaller than 45\% of the Roche lobe radius, which would make the disk
contamination completely insignificant.

We have ignored reprocessing of radiation from the secondary star, but
one can show that it is again not important.

\clearpage

\begin{deluxetable}{cccccc}
\tablecaption{Fitting Parameters for ASM Light 
Curves\tablenotemark{a}\label{ASMfits}}
\tablewidth{0pt}
\tablehead{
\colhead{Channel} &
\colhead{$a_0$} &
\colhead{$a_1$}  &
\colhead{$a_2$} &
\colhead{$\chi^2_{\nu}$} &
\colhead{$A$\tablenotemark{b}}
}
\startdata
A & $ 0.7291 \pm 0.0025$ & $-0.0263 \pm 0.0036$ & $-0.0031 \pm 0.0035$
       &   1.942  & $0.072 \pm 0.010$ \cr
B & $ 0.5275 \pm 0.0020$ & $-0.0204 \pm 0.0028$ & $ 0.0037 \pm 0.0028$
       &  1.151   & $0.077 \pm 0.011$ \cr
C & $ 0.2602 \pm 0.0026$ & $-0.0050 \pm 0.0038$ & $ 0.0005 \pm 0.0037$ 
       &  0.681   & $0.038 \pm 0.029$ \cr
SUM & $ 1.5201 \pm 0.0042$ & $-0.0491 \pm 0.0060$ & $-0.0004 \pm 0.0059$
       &  1.236   & $0.065 \pm 0.008$ \cr
\enddata
\tablenotetext{a}{$f(\phi) = a_0 + a_1\cos(2\pi\phi) + a_2\sin(2\pi\phi)$}
\tablenotetext{b}{fractional amplitude $A=(2|a_1|/a_0)$}
\end{deluxetable}

\begin{deluxetable}{cccccc}
\tablecaption{LMC X-1 Column Density Measurements\label{xray}}
\tablewidth{0pt}
\tablehead{
\colhead{Model Number} &
\colhead{Model} &
\colhead{Mission}  &
\colhead{$N_H$ ($10^{21}$cm$^{-2}$)} &
\colhead{$A_V$ (mag)} &
\colhead{Reference} 
}
\startdata
1 & DBB & {\em Chandra} HETG & $4.6\pm 0.2$ & $2.57\pm 0.12$ 
                                               & Cui et al.\ 2002 \cr
2 & COMPTT & {\em Chandra} HETG & $5.9\pm 0.3$ & $3.30\pm 0.18$ 
                                               & Cui et al.\ 2002 \cr
3 & DBB   & {\em BeppoSAX}  & $8.1\pm 0.5$ & $4.53\pm 0.29$ 
                                               & Haardt et al.\ 2001 \cr
4 & GRdisk   & {\em ASCA}  & $5.3\pm 0.2$ & $2.96\pm 0.12$ 
                                               & Gierlinski et al.\ 2001 \cr
5 & DBB   & {\em ASCA}  & $6.3\pm 0.1$ & $3.52\pm 0.08$ 
                                               & Nowak et al.\ 2001 \cr
6 & DBB   & {\em BBXRT}  & $5.8\pm 0.9$ & $3.24\pm 0.51$ 
                                               & Schlegel et al.\ 1994 \cr
7 & X-ray halo   & {\em Chandra}  & $6.5\pm 0.1$ & $3.65\pm 0.10$ 
                                               & Xiang et al.\ 2005 \cr
\enddata
\end{deluxetable}

\clearpage

\begin{deluxetable}{rccc}
\tablecaption{LMC X-1 Adopted Parameters\label{parm}}
\tablewidth{0pt}
\tablehead{
\colhead{parameter} &
\colhead{value} &
\colhead{value} & 
\colhead{Adopted}           \\
\colhead{  } &
\colhead{(circular orbit)} &
\colhead{(eccentric orbit)} &
\colhead{value\tablenotemark{a}} 
}
\startdata
$P$ (days)             
                       &  $3.90914\pm 0.00005$
                       &  $3.90917\pm 0.00005$ 
                       &  $3.90917\pm 0.00005$  \\ 
$T_0$ (HJD 2,453,300+)\tablenotemark{b} 
                       &  $91.3436\pm 0.0078$
                       &  $91.3419\pm 0.0087$  
                       &  $91.3436\pm 0.0080$ \\ 
$K_2$ (km s$^{-1}$)    
                       & $71.61\pm 0.67$
                       & $70.74\pm 0.75$  
                       & $71.61\pm 1.10$    \\ 
$i$ (deg)              
                       & $36.38\pm 1.92$ 
                       & $37.00\pm 1.78$ 
                       & $36.38\pm 2.02$    \\
$f_2$                  
                       & $0.886 \pm 0.036$ 
                       & $0.894 \pm 0.040$  
                       & $0.886 \pm 0.037$              \\
$M_2$ ($M_{\odot}$)    
                       & $31.79\pm 3.48$ 
                       & $30.62\pm 3.17$ 
                       & $31.79\pm 3.67$        \\
$e$                    
                       & \nodata
                       & $0.0256 \pm 0.0066$ 
                       & \nodata   \\
$\omega$ (deg)         
                       & \nodata
                       & $260.5  \pm 16.8$ 
                       & \nodata  \\
$T_0$ (HJD 2,453,300+)\tablenotemark{c} 
                       &  \nodata 
                       &  $91.3072\pm 0.2073$ 
                       &  \nodata \\ 
$R_2$ ($R_{\odot}$)    
                       & $16.89\pm 0.87$ 
                       & $16.41\pm 0.71$  
                       & $17.00\pm 0.80$\tablenotemark{d} \\ 
$\log g$ (cgs)         
                       & $3.485\pm 0.014$ 
                       & $3.497\pm 0.011$ 
                       & $3.485\pm 0.018$   \\
$a$ ($R_{\odot}$)      
                       & $36.49\pm 1.42$ 
                       & $35.97\pm 1.28$ 
                       & $36.49\pm 1.51$  \\
$M$ ($M_{\odot}$)      
                       & $10.91\pm 1.41$ 
                       & $10.30\pm 1.18$ 
                       & $10.91\pm 1.54$  \\
$\chi^2$ 
    ($B$, $V$ bands)   
                       & 59.18, 60.44    
                       & 59.00, 59.70    
                       & \nodata         \\
$\chi^2$ 
 (radial velocities)   
                       & 49.63           
                       & 36.20    
                       & \nodata             \\
$\chi^2$
   (total)             
                       & 169.27   
                       & 155.39
                       & \nodata
\enddata
\tablecomments{The effective temperature of the secondary has
been constrained to the range $32,200\le T_{\rm eff}\le 34,200$~K.}
\tablenotetext{a}{The uncertainties include systematic errors}
\tablenotetext{b}{Time of inferior conjunction of secondary}
\tablenotetext{c}{Time of periastron passage of secondary}
\tablenotetext{d}{Determined from the temperature, $K$ magnitude,
distance, and extinction}
\end{deluxetable}

\begin{deluxetable}{cccccc}
\tablecaption{LMC X-1 Parameters as a Function 
of $K_2$\label{fixK}}
\tablewidth{0pt}
\tablehead{
\colhead{Assumed $K_2$} &
\colhead{$i$} &
\colhead{$\Delta\phi$\tablenotemark{a}}  &
\colhead{$M_2$}  &
\colhead{$M$} &
\colhead{$\chi^2_{\rm min}$} \\
\colhead{(km s$^{-1}$)} &
\colhead{(deg)} &
\colhead{} &
\colhead{($M_{\odot}$)} & 
\colhead{($M_{\odot}$)} &
\colhead{}
}
\startdata
$68.50\pm 0.65$    &  $36.20\pm 2.02$ & $0.0111\pm 0.0055$  & $31.46\pm 3.51$ 
                          &  $10.34\pm 1.31$  &  115.68 \\ 
$69.50\pm 0.65$    &  $36.17\pm 2.02$ & $0.0111\pm 0.0057$  & $31.72\pm 3.53$ 
                          &  $10.58\pm 1.33$  &  115.66 \\ 
$69.80\pm 0.65$    &  $36.17\pm 2.03$ & $0.0111\pm 0.0055$  & $31.79\pm 3.48$ 
                          &  $10.65\pm 1.33$  &  115.65 \\ 
$70.50\pm 0.65$    &  $36.16\pm 2.03$ & $0.0111\pm 0.0055$  & $32.00\pm 3.49$ 
                          &  $10.82\pm 1.36$  &  115.63 \\ 
$71.50\pm 0.65$    &  $36.23\pm 2.01$ & $0.0112\pm 0.0055$  & $32.10\pm 3.51$ 
                          &  $11.00\pm 1.38$  &  115.60 \\ 
$72.50\pm 0.65$    &  $36.19\pm 2.02$ & $0.0112\pm 0.0055$  & $32.37\pm 3.51$ 
                          &  $11.26\pm 1.41$  &  115.57 \\ 
\enddata
\tablenotetext{a}{$\Delta\phi$ is the phase shift of the
photometric $T_0$ relative to the spectroscopically determined
value.}
\tablecomments{The assumed distance is $18.41\pm 0.10$ mag,
the assumed temperature is in the range
$32,200\le T_{\rm eff}\le 34,200$ K, and the assumed
extinction is $A_V=2.28\pm 0.06$ mag.}
\end{deluxetable}

\begin{deluxetable}{ccccccc}
\tablecaption{LMC X-1 Parameters for Nonsynchronous
Rotation\label{nonsync}}
\tablewidth{0pt}
\tablehead{
\colhead{$\Omega$\tablenotemark{a}} &
\colhead{$i$}  &
\colhead{$e$} &
\colhead{$\omega$} &
\colhead{$M_2$} &
\colhead{$M$}  &
\colhead{$\chi^2_{\rm min}$} \\
\colhead{ } &
\colhead{(deg)} &
\colhead{} &
\colhead{(deg)} &
\colhead{($M_{\odot}$)} & 
\colhead{($M_{\odot}$)} &
\colhead{}
}
\startdata
1.15       &  $32.0\pm 1.0$ & 0 (fixed) & \nodata & $30.7\pm 2.3$ 
                                            &   $12.5\pm 1.0$ & 171.53 \\ 
1.10       &  $33.0\pm 1.4$ & 0 (fixed) & \nodata & $31.4\pm 2.4$ 
                                            &   $12.1\pm 1.2$ & 170.45 \\ 
1.05       &  $34.9\pm 1.5$ & 0 (fixed) & \nodata & $31.0\pm 1.5$  
                                            &   $11.3\pm 1.4$ & 169.91 \\ 
1.00       &  $36.4\pm 1.9$ & 0 (fixed) & \nodata & $31.2\pm 3.5$  
                                            &   $10.9\pm 1.4$ & 169.38 \\ 
0.95       &  $38.5\pm 2.2$ & 0 (fixed) & \nodata & $32.2\pm 3.1$  
                                            &   $10.3\pm 1.0$ & 169.11 \\ 
0.90       &  $42.8\pm 2.5$ & 0 (fixed) & \nodata & $31.6\pm 2.9$ 
                                            &   ~$9.3\pm 1.1$ & 168.73 \\ 
0.85       &  $44.5\pm 2.6$ & 0 (fixed) & \nodata & $32.3\pm 2.9$ 
                                            &   ~$8.4\pm 0.9$ & 168.51 \\ 
0.80       &  $48.1\pm 2.6$ & 0 (fixed) & \nodata & $33.0\pm 2.2$ 
                                            &   ~$8.4\pm 1.1$ & 168.18 \\ 
0.75       &  $52.1\pm 2.5$ & 0 (fixed) & \nodata & $33.8\pm 3.1$ 
                                            &   ~$7.9\pm 0.9$ & 167.84 \\ 
0.70       &  $56.3\pm 2.7$ & 0 (fixed) & \nodata & $35.6\pm 2.9$ 
                                            &   ~$7.7\pm 0.7$ & 168.17 \\ 
0.65       &  $60.9\pm 1.8$ & 0 (fixed) & \nodata
                         & $37.4\pm 2.0$    &   ~$7.5\pm 0.4$ & 168.06 \\ 
\hline
1.15       &  $33.8\pm 1.5$ & $0.0215\pm 0.0038$ & $267.0\pm 17.2$ 
                         & $28.2\pm 2.9$    &   $11.0\pm 1.2$ & 162.45 \\ 
1.10       &  $34.6\pm 1.7$ & $0.0210\pm 0.0063$ & $265.5\pm 16.4$
                         & $29.4\pm 3.2$    &   $11.0\pm 1.3$ & 160.46 \\ 
1.05       &  $36.0\pm 1.8$ & $0.0245\pm 0.0036$ & $261.3\pm 16.8$ 
                         & $29.6\pm 3.3$    &   $10.7\pm 1.4$ & 157.03 \\ 
1.00       &  $37.4\pm 2.1$ & $0.0269\pm 0.0038$ & $266.3\pm 16.9$
                         & $30.3\pm 3.4$    &   $10.3\pm 1.4$ & 154.93 \\ 
0.95       &  $39.4\pm 2.0$ & $0.0287\pm 0.0048$ & $266.3\pm 17.2$ 
                         & $31.4\pm 3.2$    &   ~$9.8\pm 1.1$ & 153.19 \\ 
0.90       &  $42.1\pm 1.7$ & $0.0293\pm 0.0042$ & $264.5\pm 16.6$ 
                         & $32.2\pm 2.4$    &   ~$9.2\pm 0.9$ & 152.35 \\ 
0.85       &  $44.2\pm 2.2$ & $0.0311\pm 0.0046$ & $265.2\pm 17.0$ 
                         & $33.5\pm 2.8$    &   ~$9.0\pm 1.2$ & 150.76 \\ 
0.80       &  $47.6\pm 2.6$ & $0.0336\pm 0.0045$ & $266.2\pm 17.5$
                         & $34.5\pm 3.6$    &   ~$8.6\pm 1.1$ & 149.94 \\ 
0.75       &  $50.3\pm 3.5$ & $0.0360\pm 0.0038$ & $268.3\pm 17.1$ 
                         & $36.4\pm 3.5$    &   ~$8.4\pm 1.0$ & 149.48 \\ 
0.70       &  $56.8\pm 4.0$ & $0.0372\pm 0.0035$ & $268.2\pm 16.5$ 
                         & $36.1\pm 3.9$    &   ~$7.5\pm 0.9$ & 148.32 \\ 
0.65       &  $61.7\pm 2.6$ & $0.0372\pm 0.0037$ & $259.9\pm 17.0$
                         & $38.0\pm 2.4$    &   ~$7.4\pm 0.4$ & 148.46 \\
\enddata
\tablenotetext{a}{$\Omega$ is the ratio of the rotation
frequency of the star to the orbital frequency.}
\tablecomments{The temperature is in the range
$32,200 \le T_{\rm eff} \le 34,200$~K,
the assumed distance is $18.41\pm 0.10$ mag, and the assumed
extinction is $A_V=2.28\pm 0.06$ mag.}
\end{deluxetable}

\begin{deluxetable}{ccccccc}
\tablecaption{LMC X-1 Parameters as a Function 
of Temperature\label{fixT}}
\tablewidth{0pt}
\tablehead{
\colhead{temperature} &
\colhead{derived radius} &
\colhead{$\log L$}  &
\colhead{$i$}  &
\colhead{$M_2$} &
\colhead{$M$}  &
\colhead{$\chi^2_{\rm min}$} \\
\colhead{(K)} &
\colhead{($R_{\odot}$)} &
\colhead{($L_{\odot}$)} &
\colhead{(deg)} &
\colhead{($M_{\odot}$)} & 
\colhead{($M_{\odot}$)} &
\colhead{}
}
\startdata
30,000 & $17.03\pm 0.83$ & $5.33\pm 0.04$ & $36.29\pm 2.10$ & $32.34\pm 3.48$
                                            &   $11.09\pm 1.34$ & 174.14 \\ 
31,000 & $16.89\pm 0.82$ & $5.38\pm 0.04$ & $36.47\pm 2.11$ & $31.93\pm 1.6$
                                            &   $10.95\pm 1.34$ & 174.66 \\ 
32,000 & $17.07\pm 0.83$ & $5.44\pm 0.04$ & $36.12\pm 2.03$ & $32.73\pm 3.56$
                                            &   $11.21\pm 1.35$ & 173.94 \\ 
33,000 & $17.07\pm 0.82$ & $5.49\pm 0.04$ & $36.30\pm 2.02$ & $32.45\pm 3.73$
                                            &   $11.07\pm 1.27$ & 171.30 \\ 
34,000 & $16.72\pm 0.81$ & $5.53\pm 0.04$ & $37.11\pm 2.17$ & $30.67\pm 3.61$
                                            &   $10.44\pm 1.32$ & 169.24 \\ 
35,000 & $16.23\pm 0.78$ & $5.55\pm 0.04$ & $38.42\pm 2.33$ & $28.66\pm 3.21$
                                            &   ~$9.66\pm 1.25$ & 169.55 \\ 
36,000 & $15.79\pm 0.76$ & $5.58\pm 0.04$ & $39.92\pm 2.13$ & $27.12\pm 3.00$
                                            &   ~$8.96\pm 1.13$ & 170.07 \\ 
37,000 & $15.42\pm 0.74$ & $5.61\pm 0.04$ & $40.16\pm 1.71$ & $26.81\pm 1.91$
                                            &   $~8.79\pm 0.75$ & 165.82 \\ 
\enddata
\tablecomments{The assumed distance is $18.41\pm 0.10$ mag and the assumed
extinction is $A_V=2.28\pm 0.06$ mag.}
\end{deluxetable}

\begin{deluxetable}{cccccc}
\tablecaption{LMC X-1 Parameters as a Function of 
Radius\label{fixR}}
\tablewidth{0pt}
\tablehead{
\colhead{radius} &
\colhead{derived luminosity} &
\colhead{$i$}  &
\colhead{$M_2$} &
\colhead{$M$}  &
\colhead{$\chi^2_{\rm min}$} \\
\colhead{($R_{\odot}$)} &
\colhead{($\log L/L_{\odot}$)} &
\colhead{(deg)} &
\colhead{($M_{\odot}$)} & 
\colhead{($M_{\odot}$)} &
\colhead{}
}
\startdata
15.0 &  $5.39\pm 0.03$ & $41.24\pm 0.87$ & $24.56\pm 0.95$
                                            &   ~$8.15\pm 0.25$ & 168.52 \\ 
15.5 &  $5.42\pm 0.03$ & $39.64\pm 0.80$ & $26.34\pm 0.90$
                                            &   ~$8.86\pm 0.28$ & 168.71 \\ 
16.0 &  $5.45\pm 0.03$ & $38.17\pm 0.81$ & $28.17\pm 0.97$
                                            &   ~$9.62\pm 0.29$ & 168.99 \\ 
16.5 &  $5.47\pm 0.03$ & $36.82\pm 0.71$ & $30.16\pm 0.89$
                                            &   $10.43\pm 0.28$ & 169.19 \\ 
17.0 &  $5.50\pm 0.03$ & $35.62\pm 0.71$ & $32.28\pm 0.97$
                                            &   $11.29\pm 0.31$ & 169.44 \\ 
17.5 &  $5.53\pm 0.03$ & $34.42\pm 0.70$ & $34.50\pm 1.24$
                                            &   $12.22\pm 0.30$ & 169.77 \\ 
18.0 &  $5.55\pm 0.03$ & $33.37\pm 0.65$ & $36.86\pm 0.87$
                                            &   $13.20\pm 0.37$ & 170.42 \\ 
18.5 &  $5.57\pm 0.03$ & $32.39\pm 0.56$ & $39.29\pm 0.59$
                                            &   $14.22\pm 0.34$ & 170.36 \\ 
19.0 &  $5.56\pm 0.03$ & $32.48\pm 0.59$ & $42.60\pm 0.56$
                                            &   $15.44\pm 0.37$ & 171.27 \\ 
\enddata
\tablecomments{The assumed temperature is in the range
$32,200 \le T_{\rm eff} \le 34,200$~K,
the assumed distance is $18.41\pm 0.10$ mag, and the assumed
extinction is $A_V=2.28\pm 0.06$ mag.}
\end{deluxetable}
 
\clearpage

\begin{deluxetable}{ccccc}
\tablecaption{LMC X-1 Parameters as a Function of 
X-ray Heating\label{fixX}}
\tablewidth{0pt}
\tablehead{
\colhead{$\log L_x$ (erg s$^{-1}$)} &
\colhead{$i$}  &
\colhead{$M_2$} &
\colhead{$M$}  &
\colhead{$\chi^2_{\rm min}$} \\
\colhead{(isotropic equivalent)} &
\colhead{(deg)} &
\colhead{($M_{\odot}$)} & 
\colhead{($M_{\odot}$)} &
\colhead{}
}
\startdata
no heating &  $36.02\pm 1.78$ & $31.60\pm 3.11$ 
                                            &   $10.98\pm 1.28$ & 169.57 \\ 
38.00       &  $36.02\pm 1.82$ & $31.61\pm 3.03$  
                                            &   $10.99\pm 1.40$ & 169.57 \\ 
38.25       &  $36.01\pm 1.28$ & $31.54\pm 2.27$  
                                            &   $10.99\pm 0.99$ & 169.39 \\ 
38.50       &  $36.06\pm 1.83$ & $31.41\pm 3.28$ 
                                            &   $10.95\pm 1.40$ & 169.35 \\ 
38.75       &  $36.02\pm 1.39$ & $31.40\pm 2.50$  
                                            &   $10.97\pm 0.77$ & 169.48 \\ 
39.00       &  $35.91\pm 1.77$ & $31.49\pm 3.17$  
                                            &   $11.05\pm 1.16$ & 170.30 \\ 
\enddata
\tablecomments{The assumed temperature is in the range
$32,200 \le T_{\rm eff} \le 34,200$~K,
the assumed distance is $18.41\pm 0.10$ mag, and the assumed
extinction is $A_V=2.28\pm 0.06$ mag.}
\end{deluxetable}

\begin{deluxetable}{cccccc}
\tablecaption{LMC X-1 Parameters as a Function of 
the Disk Fraction\label{diskfrac}}
\tablewidth{0pt}
\tablehead{
\colhead{Disk fraction} &
\colhead{derived radius}&
\colhead{$i$}  &
\colhead{$M_2$} &
\colhead{$M$}  &
\colhead{$\chi^2_{\rm min}$} \\
\colhead{($V$-band)} &
\colhead{($R_{\odot}$)} & 
\colhead{(deg)} &
\colhead{($M_{\odot}$)} & 
\colhead{($M_{\odot}$)} &
\colhead{}
}
\startdata
0.05       & $16.58\pm 0.80$ &  $37.31\pm 1.84$ & $28.88\pm 2.56$ 
                                            &   $10.03\pm 1.10$ & 169.78 \\ 
0.10       & $16.15\pm 0.80$ &  $37.99\pm 2.28$ & $27.25\pm 2.65$  
                                            &   ~$9.52\pm 1.12$ & 170.16 \\ 
0.15       & $15.69\pm 0.80$ &  $38.44\pm 1.43$ & $25.96\pm 1.89$  
                                            &   ~$9.15\pm 0.73$ & 171.04 \\ 
0.20       & $15.22\pm 0.80$ &  $39.10\pm 1.36$ & $24.43\pm 1.70$ 
                                            &   ~$8.69\pm 0.60$ & 175.73 \\ 
\enddata
\tablecomments{The assumed temperature is in the range
$32,200 \le T_{\rm eff} \le 34,200$~K,
the assumed distance is $18.41\pm 0.10$ mag, and the assumed
extinction is $A_V=2.28\pm 0.06$ mag.}
\end{deluxetable}

\clearpage

\clearpage

\end{document}